\documentclass{aa}  
\usepackage{graphicx}
\usepackage{txfonts}
\begin{document}
   \title{Optical atmospheric extinction over Cerro Paranal\thanks{Based on 
    observations made with ESO Telescopes at Paranal Observatory.}}
  \subtitle{}
   \author{F. Patat\inst{1}
   \and
    S. Moehler\inst{1} 
    \and 
    K. O'Brien\inst{2,3}
    \and
    E. Pompei\inst{2}
    \and
    T. Bensby\inst{2,4}
    \and
    G. Carraro\inst{2,5}
    \and
    A. de Ugarte Postigo\inst{2,6}
    \and 
    A. Fox\inst{2}
    \and 
    I. Gavignaud\inst{7}
    \and
    G. James\inst{1,2}
    \and 
    H. Korhonen\inst{1}
    \and   
    C. Ledoux\inst{2}
    \and 
    S. Randall\inst{1}
    \and 
    H. Sana\inst{2,8}
    \and  
    J. Smoker\inst{2}
    \and 
    S. Stefl\inst{2}
    \and
    T. Szeifert\inst{2}   
}

   \offprints{F. Patat}

   \institute{European Organization for Astronomical Research in the 
	Southern Hemisphere (ESO), Karl-Schwarzschild-str. 2,
              85748, Garching b. M\"unchen, Germany
              \email{fpatat@eso.org}
              \and
	      European Organization for Astronomical Research in the 
	Southern Hemisphere (ESO), Alonso de C\`ordova 3107,
              Vitacura, Casilla 19001, Santiago 19, Chile 
              \and
              Department of Physics, University of California, Santa Barbara, 
              CA, USA
              \and
              Lund Observatory, Lund University, Box 43, SE-22100 Lund, Sweden
              \and
              Dipartimento di Astronomia, Universit\'a di Padova, 
              Vicolo Osservatorio 3, I-35122, Padova, Italy
              \and 
              INAF - Osservatorio Astronomico di Brera, via E. Bianchi 46, 
              23807, Merate, LC, Italy.
              \and
              Departamento de Ciencias Fisicas, Universidad Andres Bello,
              Santiago, Chile
              \and
              Universiteit van Amsterdam,
              Sterrenkundig Instituut Anton Pannekoek, Postbus 94249 - 1090 GE,
              Amsterdam, The Netherlands}

   \date{Received August 2010; accepted  ...}

\abstract{}{}{}{}{} 
 
  \abstract              
   {} 
  {The present study was conducted to
   determine the optical extinction curve for Cerro Paranal under
   typical clear-sky observing conditions, with the purpose of
   providing the community with a function to be used to correct the
   observed spectra, with an accuracy of 0.01 mag
   airmass$^{-1}$. Additionally, this work was meant to analyze the
   variability of the various components, to derive the main
   atmospheric parameters, and to set a term of reference for future
   studies, especially in view of the construction of the Extremely
   Large Telescope on the nearby Cerro Armazones.}
   {The extinction curve of Paranal was obtained through
     low-resolution spectroscopy of 8 spectrophotometric standard
     stars observed with FORS1 mounted at the 8.2 m Very Large
     Telescope, covering a spectral range 3300--8000 \AA. A total of
     600 spectra were collected on more than 40 nights
     distributed over six months, from October 2008 to March 2009. The
     average extinction curve was derived using a global fit
     algorithm, which allowed us to simultaneously combine all the
     available data.  The main atmospheric parameters were retrieved
     using the LBLRTM radiative transfer code, which was also utilised
     to study the impact of variability of the main molecular bands of
     O$_2$, O$_3$, and H$_2$O, and to estimate their column densities.}
  {In general, the extinction curve of Paranal appears to conform to
    those derived for other astronomical sites in the Atacama desert,
    like La Silla and Cerro Tololo. However, a systematic deficit with
    respect to the extinction curve derived for Cerro Tololo before
    the El Chich\'on eruption is detected below 4000 \AA. We attribute
    this downturn to a non standard aerosol composition, probably
    revealing the presence of volcanic pollutants above the Atacama
    desert. An analysis of all spectroscopic extinction curves
    obtained since 1974 shows that the aerosol composition has been
    evolving during the last 35 years. The persistence of traces of
    non meteorologic haze suggests the effect of volcanic eruptions,
    like those of El Chich\'on and Pinatubo, lasts several
    decades. The usage of the standard CTIO and La Silla extinction curves
    implemented in IRAF and MIDAS produce systematic over/under-estimates of the
    absolute flux.}
   {}


\authorrunning{F. Patat et al.}
\titlerunning{Optical atmospheric extinction over Cerro Paranal}

   \maketitle
%

\section{\label{sec:intro}Introduction}

The correction for optical atmospheric extinction is one of the
crucial steps for achieving accurate spectrophotometry from the ground
(see Burke et al. \cite{burke} for a recent review). Moreover, the
study of the extinction allows to better characterise an observing
site, enabling the detection of possible trends and effects caused by
transient events, like major volcanic eruptions. For this reason,
  many studies were conducted for a number of observatories around the
  world to derive and monitor it (T\"ug \cite{tug}; Sterken \&
  Jerzykiewics \cite{sterken77}; Guti\'errez-Moreno, Moreno \&
Cort\'es \cite{gutierrez82,gutierrez86}; King \cite{king}; Rufener
\cite{rufener}; Lockwood \& Thompson \cite{lockwood}; Angione \& de
Vaucouleurs \cite{angione}; Krisciunas et al. \cite{kevin87}; Minniti,
Clari\'a \& G\'omez \cite{minniti}; Krisciunas \cite{kevin90};
Pilachowski et al. \cite{pila}; Sterken \& Manfroid \cite{sterken};
Burki et al. \cite{burki}; Schuster \& Parrao \cite{schuster}; Parrao
\& Schuster \cite{parrao}; Kidger et al. \cite{kidger}). In the large
majority of the cases this is done by means of multi-colour, broad-
and narrow-band photometry. Spectrophotometry has been used in a
limited number of studies, mainly in connection to the calibration of
spectro-photometric standard stars (Stone \& Baldwin \cite{stone};
Baldwin \& Stone \cite{baldwin}; Hamuy et al. \cite{hamuy92,hamuy94};
Stritzinger et al. \cite{max}).

\begin{figure}
\centerline{
\includegraphics[width=9cm]{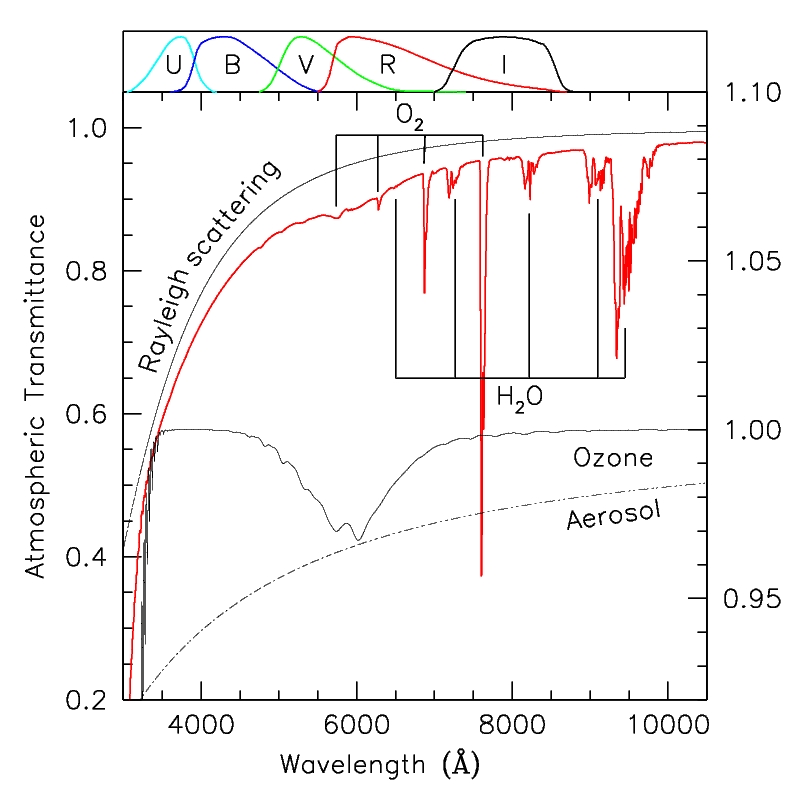}
}
\caption{\label{fig:trans} Model optical transmittance computed for
  Cerro Paranal with the LBLRTM code (see Sect.~\ref{sec:model}) for
  one airmass. The main extinction components are identified. The
  scale to the right refers to the ozone and aerosol components
  only. For presentation the model has been convolved with a Gaussian
  profile (FWHM=12 \AA). The curves on the top panel trace the
  normalised $UBVRI$ Johnson-Cousins passbands.}
\end{figure}

Although several works have been published for two astronomical sites
located in the Atacama Desert, i.e. La Silla (T\"ug \cite{tug};
  Sterken \& Jerzykiewics \cite{sterken77}; Rufener \cite{rufener};
Sterken \& Manfroid \cite{sterken}; Grothues \& Gochermann
\cite{grot}; Burki et al. \cite{burki}) and Cerro Tololo (Stone \&
Baldwin \cite{stone}; Guti\'errez et
al. \cite{gutierrez82,gutierrez86}; Stritzinger et al. \cite{max}), a
systematic work for the site of Cerro Paranal (2640 m above sea level,
24.63 degrees South), hosting the ESO Very Large Telescope (VLT), was
still missing.  Broad-band $UBVRI$ extinction coefficients are
routinely obtained at this observatory since the early phases of VLT
operations (Patat \cite{patat03}), in the framework of quality control
and instrument trending\footnote{See {\tt
    http://www.eso.org/observing/dfo/quality/}.}. In principle, as La
Silla and Cerro Tololo are placed at similar elevations above the sea
level and latitude, no significant differences are expected in the
extinction wavelength dependency, so that so far the extinction curves
for these two observatories have been widely used to correct the
optical spectra obtained at Paranal. However, Cerro Paranal is
situated much closer to the Pacific Ocean (12 km), and this might turn
into a different composition of the tropospheric aerosols, resulting
in a different extinction law.

Moreover, the extinction curves currently in use for these
observatories were derived more than 15 years ago, before the two
major eruptions of El Chich\'on (1982) and Pinatubo (1991), so that
they are most likely outdated.  The available broad-band data for
Paranal are not sufficient to address these issues. Therefore, it was
decided to undertake the PARanal Spectral Extinction Curve project
(hereafter PARSEC), with the manifold aim of a) obtaining a
spectroscopic function for the extinction correction of optical
spectra, b) getting an estimate of the variability of the various
extinction components in a time lapse free of major volcanic eruptions
affecting South America, and c) setting a term of reference for future
studies and trend analyses at this site. This is particularly
important in the view of the construction of the European Extremely
Large Telescope on Cerro Armazones, which is located only 30 km away
from Cerro Paranal. In this article we present the results of the
  PARSEC project, which was carried on between October 2008 and March
  2009.

Very briefly, the optical extinction can be described by three
separate components (Hayes \& Latham \cite{hayes}): Rayleigh
scattering by air molecules, aerosol scattering, and molecular
absorption (O$_2$, O$_3$ and H$_2$O). The contribution of the various
components is illustrated in Fig.~\ref{fig:trans}, where we present
the transmittance computed for Paranal using the LBLRTM code (Clough
et al. \cite{clough}). While the Rayleigh scattering and aerosols act
at all wavelengths, the effect of ozone in the optical is limited to
the so-called Chappuis bands (Chappuis \cite{chappuis}) between 5000
and 7000 \AA, and the Huggins bands (Huggins, \cite{huggins}) below
3400 \AA. The molecular bands of O$_2$ and H$_2$O are relevant only
above 6500 \AA.

\begin{table}
\tabcolsep 0.7mm
\caption{\label{tab:stars} PARSEC Target Stars}
\centerline{
\begin{tabular}{lccccccc}
\hline
Star  & Id. & R.A.        & DEC.      & $V$  & $B-V$ & Sp. Type & Ref. \\
      &     & (J2000)     & (J2000)   &      &       &          & \\
\hline
GD108 &   1   & 10:00:47.3 & $-$07:33:31.2 & 13.56 & $-$0.22 & sdB & 1\\
GD50  &   2   & 03:48:50.1 & $-$00:58:30.4 & 14.06 & $-$0.28 & DA2 & 1\\
EG274 &   3   & 16:23:33.8 & $-$39:13:47.5 & 11.03 & $-$0.14 & DA  & 2\\
BPM16274& 4   & 00:50:03.2 & $-$52:08:17.4 & 14.20 & $-$0.05 & DA2 & 3\\
EG21  &   5   & 03:10:31.0 & $-$68:36:02.2 & 11.38 & +0.04   & DA  & 2\\
Feige110& 6   & 23:19:58.4 & $-$05:09:55.8 & 11.82 & $-$0.30 & DOp & 1\\
GD71  &   7   & 05:52:27.5 &   +15:53:16.6 & 13.02 & $-$0.25 & DA1 & 4\\
Feige67&  8   & 12:41:51.8 &   +17:31:20.5 & 11.81 & $-$0.34 & sdO & 1\\
\hline
\multicolumn{8}{l}{(1) Oke (\cite{oke}); (2) Hamuy
et al. (\cite{hamuy92}); (3) Bohlin et al. (\cite{bohlin90})}\\
\multicolumn{8}{l}{(4) Bohlin, Colina \& Finley\cite{bohlin} }
\end{tabular}
}
\end{table}

\section{\label{sec:obs}Observations}

\subsection{Instrumental setup}

All observations were carried out using the FOcal
Reducer/low-dispersion Spectrograph (hereafter FORS1), mounted at the
Cassegrain focus of the ESO-Kueyen 8.2 m telescope (Appenzeller et
al. \cite{appenzeller}; Szeifert et al. \cite{szeifert}). At the time
of the observations discussed in this paper, FORS1 was equipped with a
mosaic of two blue-optimised, 2048$\times$4096 15$\mu$m pixel (px) E2V
CCDs. The spectra were obtained with the low-resolution G300V grism
coupled to a long, 5.0 arcsec wide slit, giving a useful spectral
range 3100--9330 \AA, and a dispersion of $\sim$3.3 \AA\/
px$^{-1}$. The spatial scale along the slit is 0.252 arcsec
px$^{-1}$. The G300V grism shows a rather pronounced second order
contamination, starting at about 6100 \AA\/ (Szeifert et
al. \cite{szeifert}). To cover the widest possible wavelength range we
run the observations using two setups, one with no filter (blue
setting) and one with the order sorting filter GG435 (red
setting). The latter provides a wavelength range free of second order
contamination 4400--8100 \AA.

The E2V detectors suffer from a marked fringing above 6500 \AA\/
(Szeifert et al. \cite{szeifert}). This becomes extremely strong for
$\lambda>$8000 \AA, reaching a peak-to-peak amplitude of $\sim$20\%,
and it is not possible to remove it from the data. Therefore, the
spectral region above this wavelength is practically unusable for our
purposes.

\begin{table}
\tabcolsep 0.4mm
\caption{\label{tab:log} Log of the observations}
\centerline{
\begin{tabular}{cccccccccc}
\multicolumn{8}{c}{GRISM 300V (blue setting)}\\
\hline
Star &   Number & Time Range     & Nights &\multicolumn{4}{c}{Airmass distribution}\\
\cline{5-8}
 Id.  &   of spectra & (MJD-54000) &    & 1.0--1.2 & 1.2--1.5 & 1.5--2.0 &  $>$2.0\\
\hline
 1 & 104 &  761.4 -- 922.0 &   28  &  35 &  17 &  36 &  16 \\ 
 2 &  76 &  743.4 -- 811.3 &   18  &  14 &  31 &  15 &  16 \\ 
 3 &  44 &  848.4 -- 921.4 &   16  &  29 &  11 &   4 &   0 \\ 
 4 &  38 &  751.1 -- 811.2 &    3  &  12 &  10 &   9 &   7 \\ 
 5 &  17 &  745.4 -- 750.4 &    6  &   0 &   0 &  17 &   0 \\ 
 6 &  11 &  751.3 -- 784.2 &    2  &   3 &   0 &   3 &   5 \\ 
 7 &  12 &  794.4 -- 805.4 &    4  &   0 &   0 &   9 &   3 \\ 
 8 &   3 &  879.4 -- 879.4 &    1  &   0 &   0 &   3 &   0 \\ 
\hline
Total & 305 &  743.4 -- 922.0 &   46  &  93 &  69 &  96 &  47 \\ 
\hline
 & & & & & & & \\
\multicolumn{8}{c}{GRISM 300V + GG435 (red setting)}\\
\hline
 1 &  92 &  761.4 -- 922.0 &   24  &  36 &  10 &  30 &  16 \\ 
 2 &  76 &  743.4 -- 811.3 &   18  &  13 &  32 &  14 &  17 \\ 
 3 &  37 &  848.4 -- 921.4 &   13  &  27 &   6 &   4 &   0 \\ 
 4 &  38 &  751.2 -- 811.2 &    3  &  12 &   8 &  11 &   7 \\ 
 5 &  19 &  745.4 -- 750.4 &    6  &   0 &   0 &  19 &   0 \\ 
 6 &  18 &  751.3 -- 784.2 &    2  &   4 &   0 &   3 &  11 \\ 
 7 &  12 &  794.4 -- 805.4 &    4  &   0 &   0 &   8 &   4 \\ 
 8 &   3 &  879.4 -- 879.4 &    1  &   0 &   0 &   3 &   0 \\ 
\hline
Total & 295 &  743.4 -- 922.0 &   43  &  92 &  56 &  92 &  55 \\ 
\hline
\end{tabular}
}
\end{table}

\subsection{Target selection and observations}

In principle, there is no need to observe spectrophotometric standards
for deriving the extinction curve, as any relatively bright star could
be used for this purpose. However, observing standard stars has the
advantage that the same data can be used also for the calibration plan
purposes, hence mitigating the impact on normal science
operations. For this reason, the programme stars were chosen among
those included in the FORS1 calibration plan. To limit the strength of
the Balmer photospheric absorption lines (which tend to hamper the
derivation of the extinction curve, especially in the blue domain),
preference was given to hot, blue objects. The selected target stars
are listed in Table~\ref{tab:stars}, which also presents their main
properties. Exposure times ranged from five seconds to a minute for
the faintest targets (i.e. GD50, BPM16274). Since the maximum shutter
timing error across the whole field of view of FORS1 is $\sim$5 ms
(Patat \& Romaniello \cite{shutter}), the photometric error associated
to the exposure time uncertainty is 0.1\% or better.

The spectroscopic data were collected on a time range spanning about 6
months, between October 4, 2008 and March 31, 2009. The target stars
were observed randomly, mainly during morning twilight. On a few
occasions the same star was observed several times during the same
night, covering a wide range in airmass (hereafter indicated as
$X$). In total, more than 300 spectra were obtained for each of the
two setups, with $X$ ranging from 1.0 to 2.6. The exact number of data
points, time and airmass ranges are shown in Table~\ref{tab:log},
together with the airmass distribution for each of the eight programme
stars. The observations were carried out under photometric or clear
conditions, which were judged at the telescope based on the
zero-points delivered by the available imaging instruments
(transparency variations $<$10\% across the whole night in the $V$
passband). The quality of the nights was also checked a posteriori,
using the data provided by the Line of Sight Sky Absorption Monitor
(LOSSAM), which is part of the Differential Image Motion Monitor
(DIMM) installed on Cerro Paranal (Sandrock et al. \cite{asm}). For
this purpose, we have retrieved the DIMM archival data for all
relevant nights, and examined the RMS fluctuation of the flux of the
star used by the instrument to derive the prevailing
seeing\footnote{The fluctuations, measured at 5000 \AA, are evaluated
  over one minute time intervals.}. When this fluctuation exceeded 2\%
(or no LOSSAM data were available), the night was classified as non
suitable, and the corresponding data rejected\footnote{The experience
  accumulated in Paranal shows that RMS fluctuations larger than 2\%
  on the timescale of minutes are most likely associated to the
  presence of clouds.}. Table~\ref{tab:log} lists only data that
passed this selection (600 out of 672 initial, non-saturated
spectra).

\begin{table}
\caption{\label{tab:seeing} FWHM image quality distribution in PARSEC
  spectra.}  \tabcolsep 5mm \centerline{
\begin{tabular}{ccccc}
\hline
$\lambda$ & $s_{min}$ & $s_{med}$ & $s_{max }$ & $s_{95}$ \\
(\AA)   & \multicolumn{4}{c}{(arcsec)} \\ 
\hline
3500 & 0.93 & 1.49 & 2.91 & 2.41 \\
4500 & 0.60 & 1.05 & 2.68 & 1.97 \\
7500 & 0.54 & 1.02 & 2.46 & 1.93 \\
\hline
\end{tabular}
}
\end{table}

All spectra were obtained with the slit oriented along the N-S
direction, which is strictly optimal only for observations close to
the meridian. The misalignment between the slit and the parallactic
angle at large hour angles (i.e. for targets at high airmass), can
potentially lead to slit losses due to the differential atmospheric
refraction (Filippenko \cite{flipper}), and the seeing
  increase at larger zenith distances (Roddier
  \cite{roddier}). As a consequence, the estimated extinction
coefficient might be systematically overestimated. To minimise this
effect we used a 5 arcsec wide slit for all observations; this,
coupled to the typical seeing attained during our observations (see
Table~\ref{tab:seeing}), ensures that differential light losses are
negligible for all data taken at $X<$2.  Additionally, FORS1 is
equipped with a Linear Atmospheric Dispersion Corrector (LADC), which
is capable of maintaining the intrinsic image quality down to airmass
$\sim$1.5 (Avila, Rupprecht \& Becker \cite{avila}). At larger
airmasses the LADC only partially reduces the effect of the
atmosphere.

The image quality values (FWHM) reported in Table~\ref{tab:seeing}
were deduced directly from the data, analysing the profiles
perpendicular to the dispersion direction at different
wavelengths. The Table presents minimum ($s_{min}$), maximum
($s_{max}$), median ($s_{med}$) and 95-th percentile ($s_{95}$) of the
observed image quality distribution. The image quality turns out to be
better than $\sim$1.5 arcsec in 50\% of the cases.  While we did not
attempt to account for the effects of atmospheric refraction, we have
applied a correction for the slit losses caused by seeing. The method
is described in Appendix~\ref{sec:seeing}.

\section{\label{sec:redu}Data reduction}

The data were processed using the FORS Pipeline (Izzo, de Bilbao \&
Larsen \cite{izzo}). The reduction steps include de-bias, flat-field
correction, and 2D wavelength calibration. The spectra were then
extracted non-interactively using the {\tt apall} task in
IRAF\footnote{IRAF is distributed by the National Optical Astronomy
  Observatories, which are operated by the Association of Universities
  for Research in Astronomy, under contract with the National Science
  Foundation.} after optimizing the extraction parameters.  Because of
the wide (5 arcsec) slit used to minimise flux losses, the uncertainty
in the target positioning within the spectrograph entrance window is
expected to produce significant shifts in the wavelength solution. For
this reason, for each standard star we selected a low-airmass template
spectrum (drawn from the data sample), to which we applied a rigid
shift in order to match O$_2$ and H$_2$O atmospheric absorption
bands\footnote{The exposures were too shallow to show night sky
  emission lines.}. Then, the shift to be applied to each input
spectrum was computed via cross correlation to the corresponding
template spectrum. For doing this we first subtracted the smooth
stellar continuum estimated by the {\tt continuum} task in
IRAF. Wavelength shifts exceeding 10 \AA\/ ($\sim$3 px) were recorded
during this process. 

The RMS error of the wavelength solution is of the order of
1\AA. Since the bluest line used by the FORS1 pipeline is \ion{He}{i}
3888\AA, systematic deviations at shorter wavelengths could in
principle occur. To check this possibility, we inspected the
calibration frames after applying to them the same wavelength solution
used for the programme stars. Thanks to the enhanced blue sensitivity
of the EEV detector, several lines (identified as \ion{Hg}{i} 3650
\AA, \ion{Cd}{i} 3610 \AA\/ and \ion{Cd}{i} 3466, 3468\AA), are
clearly detectable. The measured wavelengths of these features turn
out to be within $\sim$3\AA\/ from their laboratory values,
  corresponding to $\sim$1/4 of the typical FWHM resolution. This
indicates that the wavelength solution is accurate to within a few
\AA\/ down to $\sim$3400 \AA. To quantify the effect of
  wavelength calibration inaccuracies of this order, we derived the
  extinction curve applying a wavelength offset to the data. For an
  offset of $\pm$3.3 \AA\/ (one pixel), the variation is below
  $\mp$0.001 mag airmass$^{-1}$ in the red, while it reaches
  $\mp$0.003 mag airmass$^{-1}$ at the blue edge ($\sim$3400 \AA).

Since the adopted slit width is much wider than the typical seeing
characterizing our data set, the actual resolution in the spectra
depends on the prevailing seeing conditions.  While this has no
appreciable influence on the smooth stellar continuum, Balmer lines
are expected to be affected, especially in their cores. In principle,
this can be mitigated by matching the resolutions of the input spectra
by a Gaussian profile convolution. However, in sight of the targeted
resolution of the extinction curve, which is several tens of \AA\/
(see Sect.~\ref{sec:ext}), and the fact that different stars have
distinct line profiles and depths anyway, we have not attempted to
correct for this effect. As a consequence, weak spurious features are
expected at the wavelengths corresponding to the strongest hydrogen
lines, such as H$\alpha$ and H$\beta$.

During the PARSEC campaign, no instrument intervention or mirror
re-aluminization has occurred, guaranteeing the homogeneity of the
data. However, it is known that the main mirrors of the VLT are
subject to a reflectivity loss, mainly due to aluminum oxidation. The
losses are as large as $\sim$13\% per year in the $U$ band, they
decrease at redder wavelengths, and reach $\sim$5\% per year in the
$I$ band (Patat \cite{patat03}). Since the data presented in this
paper span over six months, the implied efficiency variation is
significantly higher than the measurement errors, and this is expected
to increase the noise. To account for this effect, at least to a first
order, the measured instrumental magnitudes were corrected using the
time elapsed from the previous aluminization (May 17, 2008;
MJD=54603.5), and a linear interpolation of the values deduced from
broad band photometry (Patat \cite{patat03}) to the required
wavelengths. The application of this correction reduces the RMS
scatter from the best fit solution by $\sim$25\%.

Finally, airmasses were derived from the unrefracted zenith distances
by means of the classical formula proposed by Hardie
(\cite{hardie}). We note that since $\sim$83\% of the spectra were
obtained at $X\leq$2, airmasses computed using other formulations
discussed in the literature are always to within $\sim$0.001.

\begin{figure}
\centerline{
\includegraphics[width=9cm]{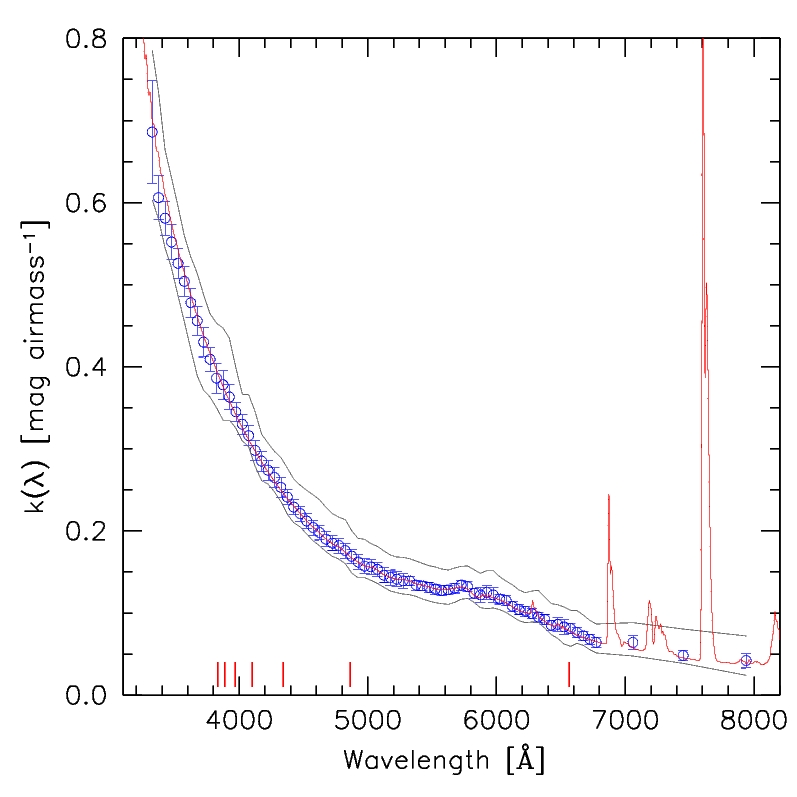}
}
\caption{\label{fig:ext} Best fit extinction coefficient for the blue
  (circles) and red (triangles) settings. The errorbars indicate the
  3-sigma uncertainty of the best fit solution. The upper and lower
  grey curves trace the 95\% variation level of single
  measurements. The solid curve is an LBLRTM simulation for Paranal
  (see Sect.~\ref{sec:model}), while the vertical segments at the
  bottom mark the positions of the first 7 Balmer lines.}
\end{figure}

\section{\label{sec:ext}Derivation of the extinction curve}

The first step in the derivation of the extinction curve is the
calculation of instrumental magnitude within predefined spectral
bins. This is defined as follows:

\begin{displaymath}
m(\lambda) = -2.5 \log \int_{\lambda-\Delta \lambda/2}^{\lambda+\Delta \lambda/2} 
s(\lambda) \; d\lambda
\end{displaymath}

\noindent where $s(\lambda)$ is the observed spectrum (in electrons
s$^{-1}$) and $\Delta \lambda$ is the spectral bin size. The fluxes
within the bins were computed by numerical integration of the
extracted, wavelength-calibrated spectra.  In the following we will
always consider bins of $\Delta \lambda$=50 \AA\/ (15 px). This choice
is motivated by three reasons: a) the resulting resolution is
sufficient to follow in great detail the behavior of the continuum
extinction; b) since the bin width is significantly larger than the
typical spectral resolution ($\sim$15 \AA\/ FWHM), the values measured
within adjacent bins should be completely uncorrelated, and c) the
resulting nominal signal-to-noise ratio per resolution element ranges
from 100 to 1200 across the whole wavelength range covered by the
observations (95\% of the bins have a signal-to-noise ratio larger
than 500). Since below $\sim$7000 \AA\/ all spectra are dominated by
  the object's shot-noise, the precision of the instrumental
  magnitudes is of the order of 1\% or better. Above this wavelength
  the dominant source of uncertainty is fringing, which reaches
  peak-to-peak amplitudes of $\sim$20\%. This limits the usable range
  to wavelengths shorter than 8100 \AA. The instrumental
    magnitudes were computed within adjacent bins up to 6800 \AA. To
    mitigate the effect of fringing, at larger wavelengths we have
    selected three bins (having widths of 160, 200, and 200 \AA),
    centred at 7060, 7450, and 7940 \AA\/ respectively, in order to
    avoid the O$_2$ and H$_2$O bands (Fig.~\ref{fig:trans}).

Once the instrumental magnitudes are corrected for the efficiency
degradation (see previous section) and slit losses due to bad seeing
(see Appendix~\ref{sec:seeing}), they can be used to finally derive
the extinction curve.

Among all possible algorithms for combining data obtained for
  different programme stars, we opted for the global solution proposed
  by Hayes \& Schmidtke (\cite{hs}). This method is applicable to
spectro-photometry and narrow-band photometry, i.e. when the passbands
are nearly monochromatic, as is our case. Under these circumstances
there are no colour-terms to be taken into account, and each
wavelength bin can be treated independently from the others. One issue
with this method is that, by construction, one cannot also solve for
the zero-point of the linear relation at the same time that the slope
(i.e. the extinction coefficient) is being determined. However, for
our purposes this is not a problem, since we are only interested in
the extinction term.

The result is presented in Fig.~\ref{fig:ext}. As the values
  derived from the two settings in the intersection region (4400--6075
  \AA) are fully consistent within the estimated uncertainties, we
  averaged them. Also, we replaced the values corresponding to the
  strong Balmer lines with a linear interpolation between the adjacent
  bins. In Fig.~\ref{fig:ext} we also included the 95\% confidence
  level deduced from the distribution of $k(\lambda)$ derived from
  each single observation (the extinction variations are discussed in
  more detail in Sect.~\ref{sec:var}).

\section{\label{sec:model}Comparison with atmospheric model}

To derive the basic physical parameters related to the extinction
curve of Paranal we used the Line By Line Radiation Transfer Model
(LBLRTM; Clough et al. \cite{clough}). This code\footnote{See {\tt
    http://rtweb.aer.com/lblrtm.html}.}, based on the HITRAN
database (Rothman et al. \cite{rothman}), has been validated against
real spectra from the UV to the sub-millimeter, and is widely used for
the retrieval of atmospheric constituents. LBLRTM solves the radiative
transfer using an input atmospheric profile, which contains the height
profiles for temperature, pressure, and chemical composition. The code
also includes the treatment of continuum scattering, and has an
internal model for tropospheric aerosols (based on LOWTRAN).

\begin{figure}
\centerline{
\includegraphics[width=9cm]{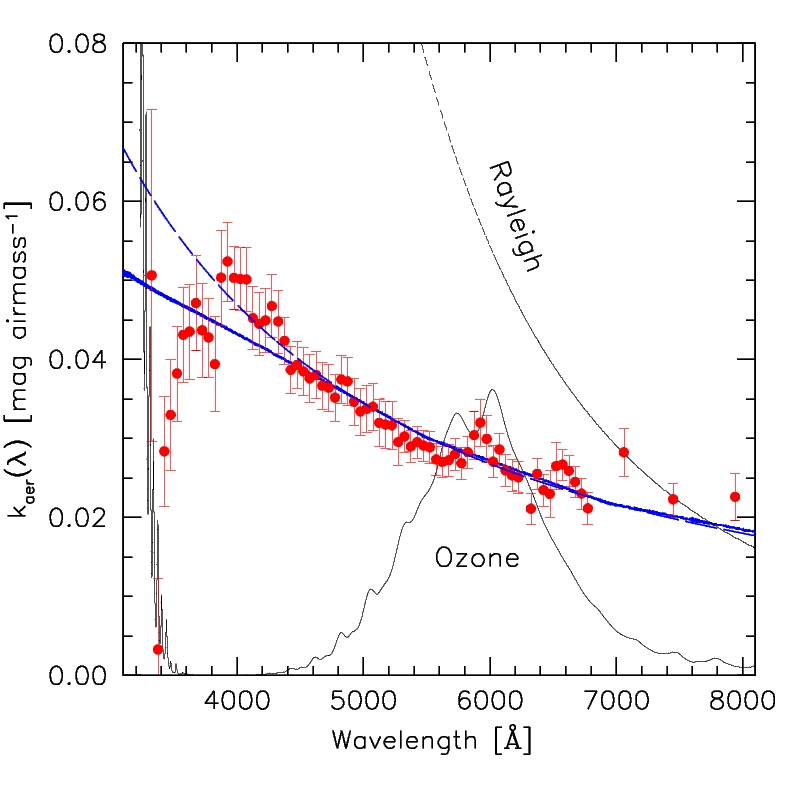}
}
\caption{\label{fig:aer} Aerosol extinction derived from the observed
  data and the LBLRTM model. The long-dashed line is a best fit model
  for $k_{\rm{aer}}(\lambda)$, with $k_0$=0.014 and $\alpha$=$-$1.42
  (see text). Overplotted are also the Rayleigh (short-dashed),
    ozone (solid thin curve), and clean tropospheric aerosol (thick
    solid) components computed by LBLRTM (see text).}
\end{figure}

For the atmospheric profiles we adopted a standard equatorial profile
as a basis for all calculations. To make the simulations more
realistic we then replaced the standard profiles of pressure,
temperature, and water content with an average vertical profile
derived from the Global Data Assimilation System\footnote{GDAS is
  maintained by the Air Resources Laboratory ({\tt
    http://www.arl.noaa.gov/}).} (GDAS). The average profile, provided
to us by W. Kausch and M. Barden (Kausch \& Barden 2010, private
communication), was obtained averaging GDAS data over the last 6 years
for the Paranal site. The corresponding amount of precipitable water
vapor (PWV) is 2.0 mm, while the ozone column density is 238.8 Dobson
Units (DU)\footnote{One DU corresponds to an ozone column density of
  2.69$\times$10$^{16}$ molecules cm$^{-2}$.}. Vacuum wavelengths were
converted into air wavelengths using the relation derived by Morton
(\cite{morton}).

\begin{figure}
\centerline{
\includegraphics[width=9cm]{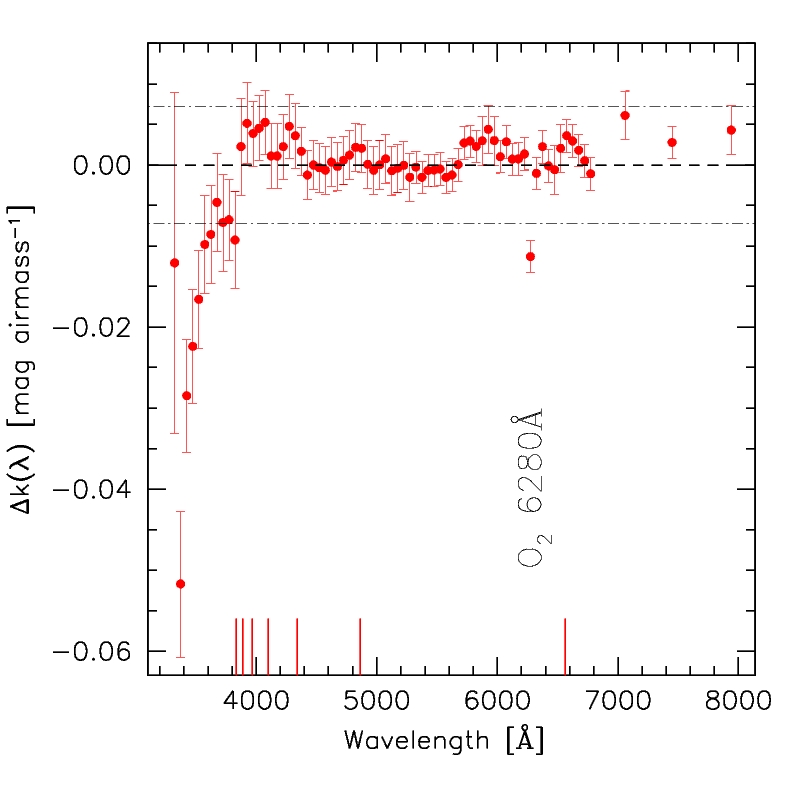}
}
\caption{\label{fig:resid} Deviations of the derived extinction curve
  from the LBLRTM simulation for Cerro Paranal (observed minus
  model). Errorbars are at the 1-sigma level. The deviation seen at
  about 6300 \AA\/ corresponds to the weak O$_2$ absorption peaking at
  $\sim$6280 \AA. The dotted-dashed lines indicate the 3-sigma
  deviation from the best fit solution in the range 4000--6800 \AA\/
  ($-$0.007 mag airmass$^{-1}$). The vertical segments at the bottom
  mark the positions of the first 7 Balmer lines.}
\end{figure}

Following Hayes \& Latham (\cite{hayes}), rather than assuming an
aerosol model, we derived it from the data. For doing this we disabled
the aerosol calculation in LBLRTM, and we computed the aerosol term
$k_{\rm{aer}}(\lambda)$ as the difference between the data and the
resulting model. The outcome is displayed in Fig.~\ref{fig:aer}. As
first suggested by \r{A}ngstrom (\cite{angstrom29,angstrom64}), the
aerosol extinction can be described as $k_{\rm{aer}}(\lambda)=k_o
\lambda^{-\alpha}$, where $k_0$ and $\alpha$ depend on the column
density and size distribution of the aerosol mixture. This analytical
formulation has been used in a number of extinction studies (Hayes \&
Latham \cite{hayes}; Angione \& de Vaucouleurs \cite{angione};
Guti\'errez-Moreno et al. \cite{gutierrez82,gutierrez86}; Rufener
\cite{rufener}; Minniti et al. \cite{minniti}; Burki et
al. \cite{burki}; Mohan et al. \cite{mohan}; Schuster \& Parrao
\cite{schuster}), and the typical value for $\alpha$ is $-$1.4$\pm$0.2
     (with $\lambda$ expressed in $\mu$m; see for instance Burki
     et al. \cite{burki}). This value is common for the so-called
     meteorological haze (i.e. free of volcanic pollutants) which
     characterises photometrically stable nights.

A reasonable fit to our data {\bf above 4000 \AA\/} is achieved for
$k_0$=0.013$\pm$0.002 mag airmass$^{-1}$ and
$\alpha=-$1.38$\pm$0.06. This result is fully in line with those found
for other astronomical sites in the Atacama desert, like Cerro Tololo
(Guti\'errez et al. \cite{gutierrez82}), and La Silla (Rufener
\cite{rufener}; Burki et al. \cite{burki}). However, the down-turn
visible below 4000 \AA\/ is probably indicating a departure from a
simple power-law model. We will get back to this issue while
discussing the comparison with other observatories in the Atacama
desert (Sect.~\ref{sec:dis}). Although the properties of aerosols and
the local conditions are difficult to model (Hess, Koepke \& Schult
\cite{hess}), for the sake of completeness we plotted the LBLRTM clean
tropospheric aerosol model in Fig.~\ref{fig:aer} (solid thick
curve). In order to reach a reasonable fit we had to reduce by 25\%
the LBLRTM default aerosol column density in the first 10 km. Even
larger reductions are necessary if one uses the marine or desert
aerosol types implemented in the code. 

To obtain information on the aerosol characteristics, we have used the
Optical Properties of Aerosol and Clouds (OPAC) software package (Hess
et al. \cite{hess}). A best fit to the data above 4000\AA\/ indicates
that the mixture is mostly composed by water soluble particles, with
very low contamination by insoluble material, minerals and sea salt.

The final model for Paranal, obtained adding the best fit aerosol term
to the LBLRTM results for the Rayleigh and ozone contributions, is
plotted in Fig.~\ref{fig:ext}, where it is compared to the data. The
model reproduces very well the overall shape, and it also accurately
follows the bumps in the 5800--6000 \AA\/ region related to the ozone
component and O$_2$ (see Fig.~\ref{fig:trans}).  The quality of the
match is better seen in Fig.~\ref{fig:resid}, where we plot the
residuals: the relative deviations are less than 0.01 mag
airmass$^{-1}$ for the vast majority of the data. The largest
discrepancies are seen redwards of 7000 \AA, and bluewards of
$\sim$4000 \AA\/. While those in the red region are most likely
explained by fringing, the deviations seen in the blue are more
difficult to explain. Although most of the data points are formally
consistent with a null deviation within a 3-sigma level, there is a
clear trend which is suggestive of a systematic effect (see below, and
the discussion in Sect.~\ref{sec:dis}).

\begin{figure}
\centerline{
\includegraphics[width=9cm]{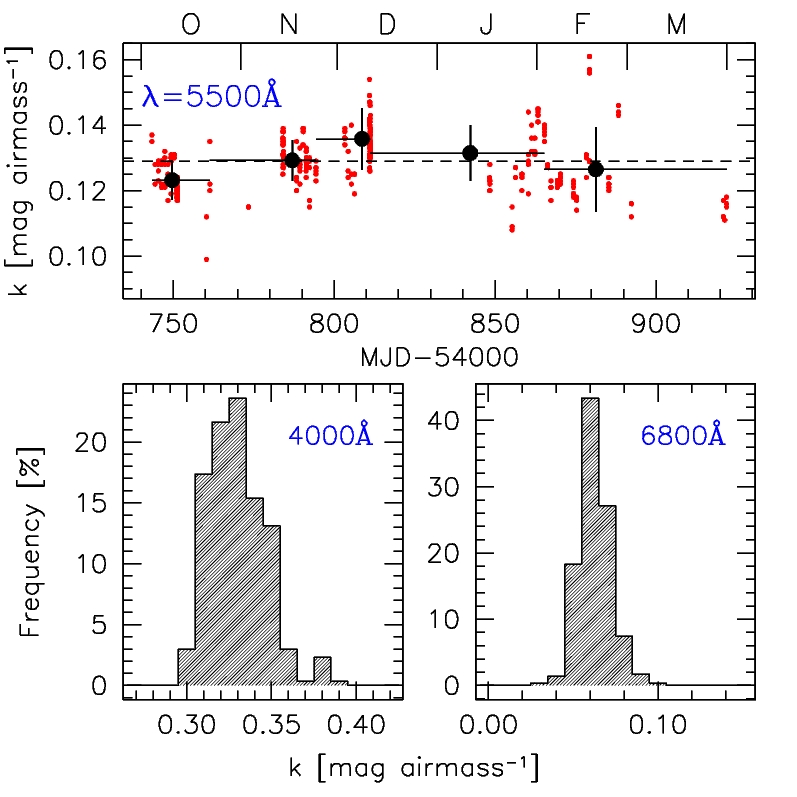}
}
\caption{\label{fig:timetrend} Upper panel: time evolution of the
  extinction at 5500 \AA. The horizontal dashed line marks the best
  fit value. The large dots are the average values computed within the
  time intervals indicated by the horizontal bars.  The vertical bars
  are the associated RMS deviations The time intervals include the
  same number of measurements (61), and the points are placed at the
  average time within each bin. Lower panel: distribution of the
  extinction coefficient at 4000 \AA\/ (left) and 6800 \AA\/ (right).}
\end{figure}

As far as the ozone content is concerned, we notice that the good fit
achieved in the region of interest (5000 to 7000 \AA) indicates this
is, on average, as predicted {by LBLRTM} for the site location,
i.e. $\sim$240 DU.

\section{\label{sec:var} Extinction variability}

An example of the extinction time evolution is presented in the upper
panel of Fig.~\ref{fig:timetrend} for $\lambda$=6000 \AA. Although
there might be an hint of a mild modulation in the observed values
(see the large dots in Fig.~\ref{fig:timetrend} and the discussion in
Sect.~\ref{sec:dis}), the limited time coverage of our sample ($\sim$6
months) does not allow us to firmly derive possible seasonal trends.
Nonetheless, the available data are sufficient to study the extinction
fluctuations under what can be considered as typical clear-sky
conditions on Paranal.

\subsection{Variability of continuum extinction}

In general, the distribution of the extinction coefficient
appears to be slightly skewed towards larger values, as found by
several authors (see for instance Reimann, Ossenkopf \& Beyersdorfer
\cite{reimann}; Kidger et al.  \cite{kidger}; Parrao \& Schuster
\cite{parrao}). This is illustrated in the lower panel of
Fig.~\ref{fig:timetrend}, where we present the distributions for two
different wavelengths (4000 and 6800 \AA). The distribution appears to
be broader in the blue than in the red. The peak-to-peak variations at
the 95\% level show a minimum value of 0.04 mag airmass$^{-1}$ at
about 6000 \AA, and grow to $\sim$0.07 mag airmass$^{-1}$ at 4000
\AA. The inter-quartile range is $\sim$0.01 mag airmass$^{-1}$ for
$\lambda>$4000 \AA, while it reaches 0.04 mag airmass$^{-1}$ at the
blue edge of the spectral range.

\begin{figure}
\centerline{
\includegraphics[width=9.6cm]{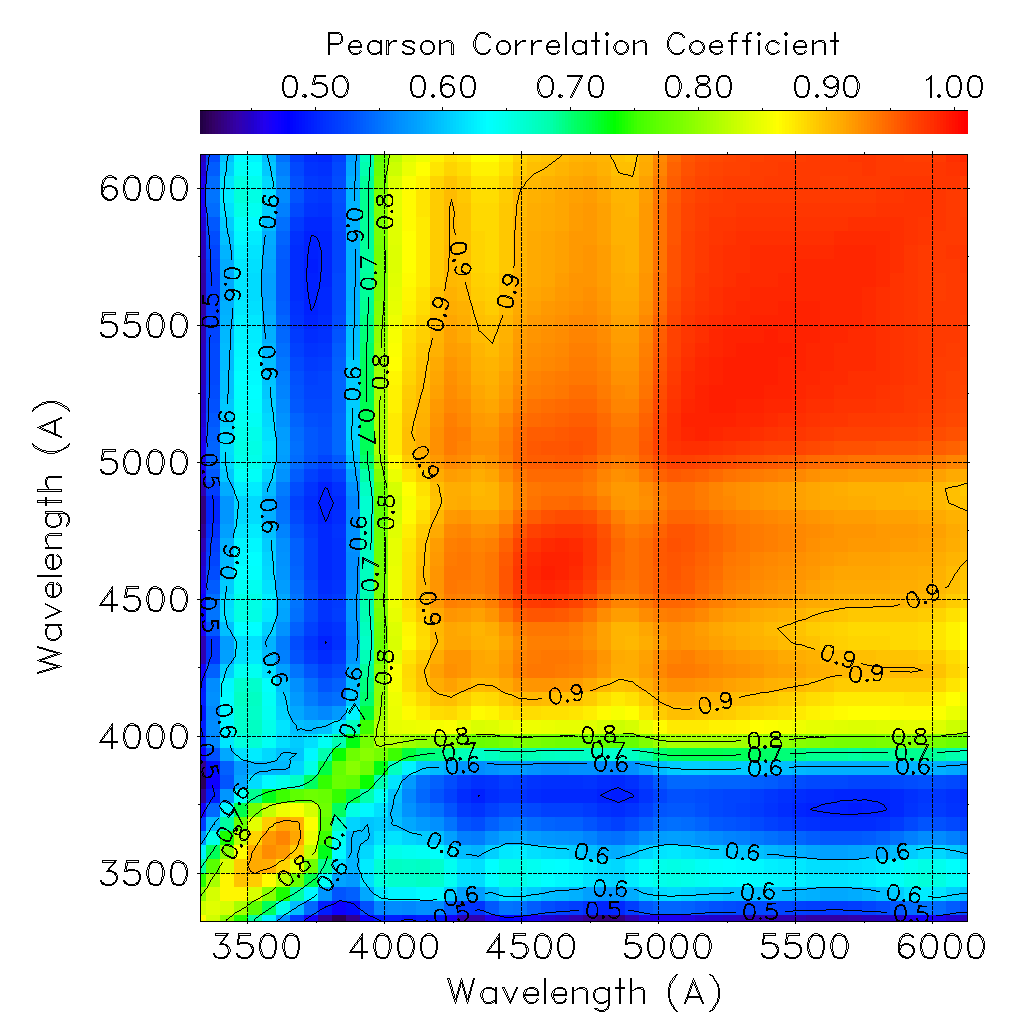}
}
\caption{\label{fig:corrmapblu}Extinction correlation map for the blue
  setting. The contours trace the iso-correlation levels. For
  presentation the map has been smoothed with a Gaussian filter
  ($\sigma$=75\AA\/ in both directions). The vertical and horizontal
  strips correspond to the Balmer lines. The decrease of correlation
  seen along the diagonal between 3700 and 4000 \AA\/ is due to the
  smoothing.}
\end{figure}

An important aspect is that the errorbars at various wavelengths are
much larger than the scatter shown by the data points along the
extinction curve (see Fig.~\ref{fig:aer}). The explanation is that the
estimated errorbars are dominated by overall variations of the
extinction rather than by random errors.  In order to show this
behavior in a more quantitative way, we studied the correlation
between the extinction measured within the different wavelength
bins. Since, for any given spectrum, $k(\lambda)$ is measured exactly
under the same conditions, this allows us to characterise the relation
between the extinction variations across the whole spectral range
covered by each of our setups separately. The result of this operation
for the blue setting is presented in Fig.~\ref{fig:corrmapblu}, which
has been calculated using the Pearson coefficient $r_{xy}$ as a figure
of merit for the correlation.

Two important facts emerge from this analysis. The first is that the
correlation is strong ($r_{xy}>$0.8) above 4000 \AA, implying that the
various portions of the extinction curve tend to vary in unison, as
expected in the case of changes in the aerosol
content/composition. The second is that the extinction at
$\lambda<$4000 \AA\/ appears to be less tightly correlated
($r_{xy}<$0.7) to that measured at larger wavelengths. We believe this
is only partially due to the larger random errors, and it is caused by
an independent behavior of the extinction curve below 4000 \AA\/ (see
the discussion in Sect.~\ref{sec:dis}).  Although the correlation is
always very significant above 4000 \AA, it becomes weaker for
$\lambda>$5700 \AA, in the sense that the extinction in the red is
less correlated with that measured below 5500 \AA. One possible
explanation for this is a combined effect of the independent
variations in the aerosol (which affects all wavelengths), and in the
ozone content (which mostly impacts the region between 5000 and 7000
\AA).

\begin{figure}
\centerline{
\includegraphics[width=9cm]{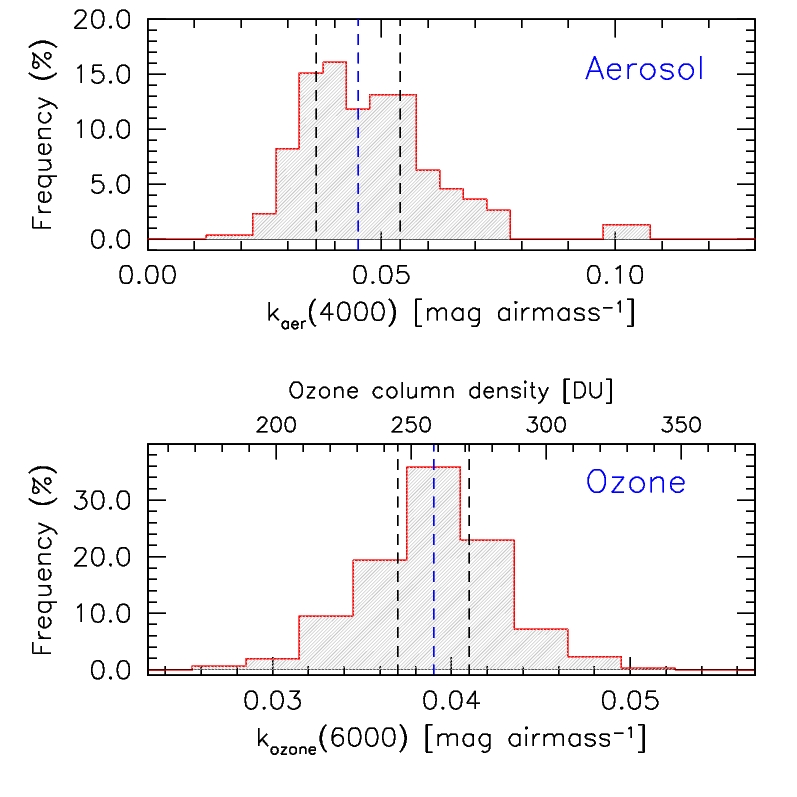}
}
\caption{\label{fig:aervar} Upper panel: distribution of aerosol
  extinction at 4000 \AA. Lower panel: distribution of ozone extinction
  at 6000 \AA. The upper scale indicates the corresponding ozone column
  density in DU (see text). The vertical dashed lines indicate the
  median, first and third quartile of the distributions.}
\end{figure}

\subsubsection{\label{sec:ray} Rayleigh scattering}

During the nights used for the PARSEC project, the atmospheric
pressure (retrieved from the Ambient Monitor, Sandrock et
al. \cite{asm}) showed a peak-to-peak excursion of 6.5 mb around the
average value of 743.6 mb (RMS variation is 1.1 mb). Since the
relative variation of the Rayleigh term is proportional to the
relative variation of atmospheric pressure (see for instance Hansen \&
Travis \cite{hansen}), this is expected to be constant to within
0.15\% (RMS), with peak-to-peak fluctuations of less than
1\%. Therefore, larger extinction variations can be attributed to
changes in the aerosol content and, to a smaller extent and only
within the wavelength range 5000--7000 \AA, to variations in the
stratospheric ozone column density (see Sect.~\ref{sec:ozone}).

\subsubsection{\label{sec:aer} Aerosol}

To estimate the variability of the aerosol component, we have computed
the residual extinction at 4000 \AA\/ (where the ozone contribution can
be neglected), after subtracting the LBLRTM Rayleigh term to the blue
setting data. The resulting distribution of the aerosol extinction,
which we indicate as $k_{\rm{aer}}(4000)$, is presented in
Fig.~\ref{fig:aervar} (upper panel). The median value is 0.045 mag
airmass$^{-1}$, and the semi-interquartile range is 0.009 mag
airmass$^{-1}$. Only in $\sim$1\% of the cases $k_{\rm{aer}}$ is smaller
than 0.01 mag airmass$^{-1}$, while it can reach values larger than
0.10 mag airmass$^{-1}$ ($\sim$2.5\%).

\subsection{\label{sec:bands} Variability of molecular bands}

In the wavelength range covered by our observations, the most relevant
absorption bands are from molecular oxygen (O$_2$ and O$_3$) and
water. While the Chappuis bands of O$_3$ are important between 5000
and 7000 \AA, O$_2$ and water mostly affect wavelengths larger than
6500 \AA\/ (see Fig.~\ref{fig:trans}). In the following we briefly
discuss the effect of their variability.

\begin{figure}
\centerline{
\includegraphics[width=9cm]{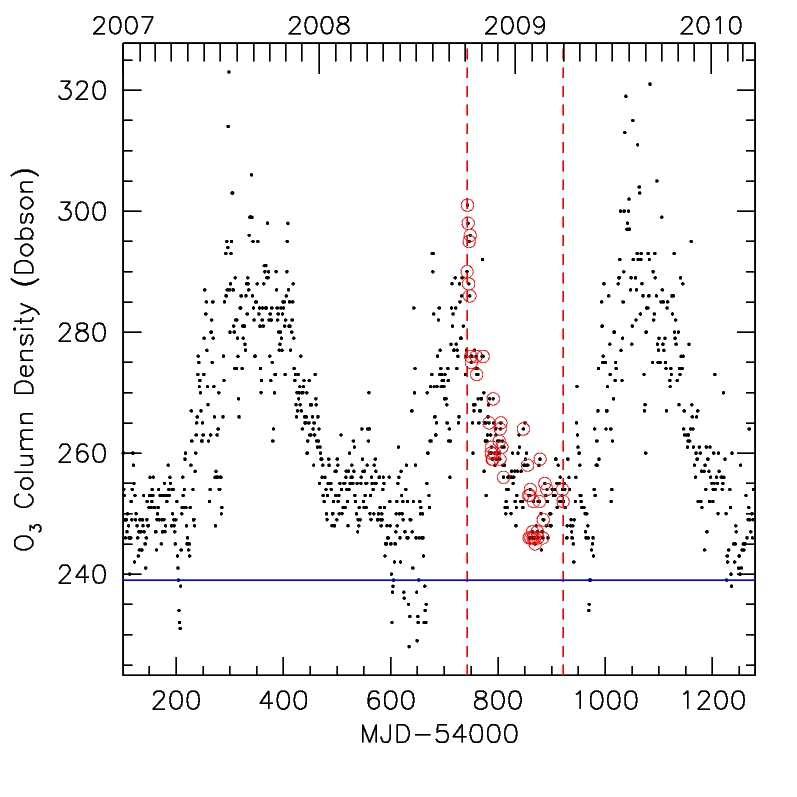}
}
\caption{\label{fig:ozone} Ozone column density for the Paranal site
  measured by the Ozone Monitor Instrument on board of AURA. The two
  vertical dashed lines indicate the time range covered by PARSEC
  data. The circles mark the dates when PARSEC data were obtained.}
\end{figure}

\subsubsection{\label{sec:ozone}Ozone bands}

Ozone is the main responsible for the cutoff in the atmospheric
transmittance below 3400 \AA, where the O$_3$ component quickly
surpasses the Rayleigh scattering (the ozone term reaches $\sim$2.5
mag airmass$^{-1}$ at 3000 \AA). Since our data do no cover this
wavelength region, we estimated the ozone column density measuring the
extinction $k_{\rm{O}_3}$ corresponding to the peak of the Chappuis
bands at $\sim$6000 \AA\/ (see Fig.~\ref{fig:aer}). This was achieved
removing the Rayleigh contribution (which was assumed to be constant
in time), and an estimate of $k_{\rm{aer}}$ at 6000 \AA. Since the
aerosol is strongly variable, for each data set we have extrapolated
the value measured at 4000 \AA\/ (cf. Sect.~\ref{sec:aer}),
conservatively assuming that $k_{\rm{aer}}$ follows the usual power
law with $\alpha$=$-$1.38, so that $k_{\rm{aer}}(6000)\approx 0.6\;
k_{\rm{aer}}(4000)$. The distribution of $k_{\rm{O}_3}$ is shown in
Fig.~\ref{fig:aervar} (lower panel). The average value is 0.039 mag
airmass$^{-1}$ (0.002 mag airmass$^{-1}$ RMS), which is slightly
larger than the LBLRTM prediction for Paranal (0.036 mag
airmass$^{-1}$). The peak-to-peak variation is about 0.02 mag
airmass$^{-1}$, and this would be sufficient to produce an
enhanced extinction variability, at least at the Chappuis bands peak
($\sim$6000 \AA). However, a significant fraction of the variance seen
in the ozone extinction is most likely due to measurement errors, and
uncertainties in the estimate of the underlying aerosol contribution.

\begin{figure}
\centerline{
\includegraphics[width=9cm]{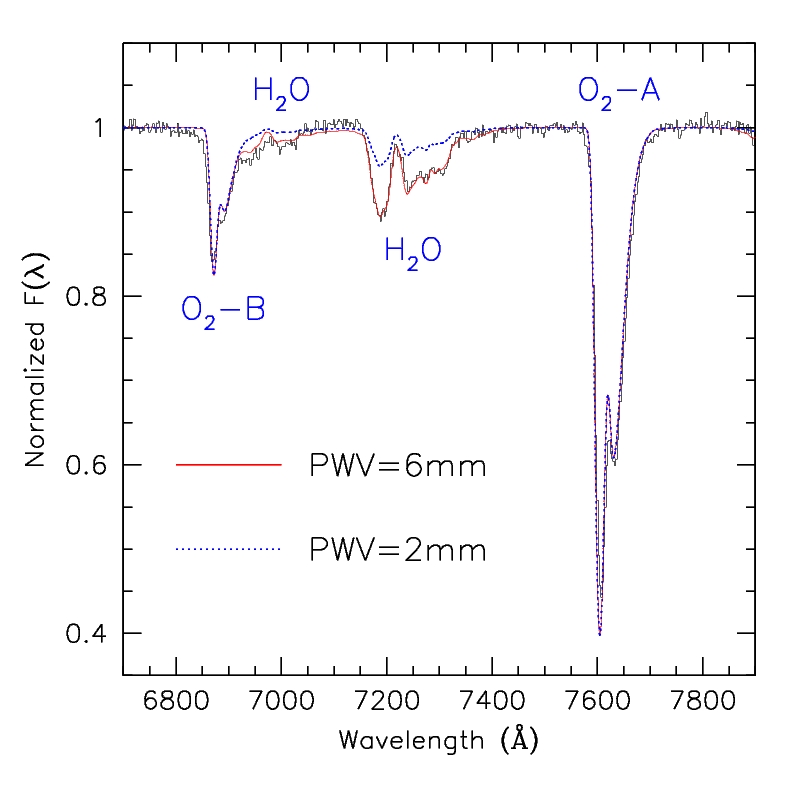}
}
\caption{\label{fig:h2o} Example spectrum of Feige~110 obtained with
  FORS1 on March 13, 2006 (grism G300I+OG590; airmass $X$=1.03). The
  signal-to-noise ratio per pixel on the continuum is $\sim$200. The
  main molecular absorption bands are marked. Superimposed are two
  LBLRTM simulations with PWV=2mm (dotted curve), and PWV=6mm (solid
  curve). In both cases the O$_2$ column density is
  3.3$\times$10$^{24}$ cm$^{-2}$. Note that at the time this spectrum
  was obtained, FORS1 was equipped with a Tektronik detector much less
  affected by fringing.}
\end{figure}

To retrieve the implied ozone column density N(O$_3$), we have run a
series of LBLRTM simulations scaling the O$_3$ profile to obtain a
total N(O$_3$) between 60 and 480 DU. In this range $k_{\rm{O}_3}(6000)$
scales linearly with the column density, and a best fit to the model
data gives the following relation:

\begin{displaymath}
k_{\rm{O}_3}(6000) \simeq 1.51 \times 10^{-4} \; \mbox{N(O$_3$)}
\end{displaymath}

\noindent where $k_{\rm{O}_3}(6000)$ is expressed in mag airmass$^{-1}$,
and N(O$_3$) is expressed in DU. From this we derive an average O$_3$
column density of 258 DU (the RMS deviation is 14 DU).

To cross-check this result, we run an independent analysis of the
ozone variability using the satellite data provided by the Ozone
Monitor Instrument (OMI) on board of NASA AURA\footnote{Data can be
  obtained from {\tt toms.gsfc.nasa.gov.}}. OMI derives the daytime
ozone column density by comparing the amount of back-scattered solar
radiation in the UV and in the optical.  The OMI O$_3$ column density
over Paranal collected between 2007 and 2009 clearly shows a seasonal
trend (see Fig.~\ref{fig:ozone}), with maxima attained around August,
September and October ($\sim$300 DU), and minima reached in February,
March and April ($\sim$240 DU). The average value during the PARSEC
campaign was about 260 DU, which is slightly larger than the value
given by the LBLRTM model ($\sim$240 DU; see previous section). A
closer inspection to the OMI data shows that the ozone content
steadily decreased during the time covered by our observations. The
peak-to-peak variation is about 25\%. This turns into a maximum
variation of $\sim$0.01 mag airmass$^{-1}$ at $\sim$6000 \AA\/ during
the time covered by our observations. An inspection of the time
evolution of $k_{\rm{O}_3}(6000)$ shows no traces of such a steady
decrease, and the fluctuations appear to be dominated by short
timescale variations, partially attributable to random errors.

\begin{figure}
\centerline{
\includegraphics[width=9cm]{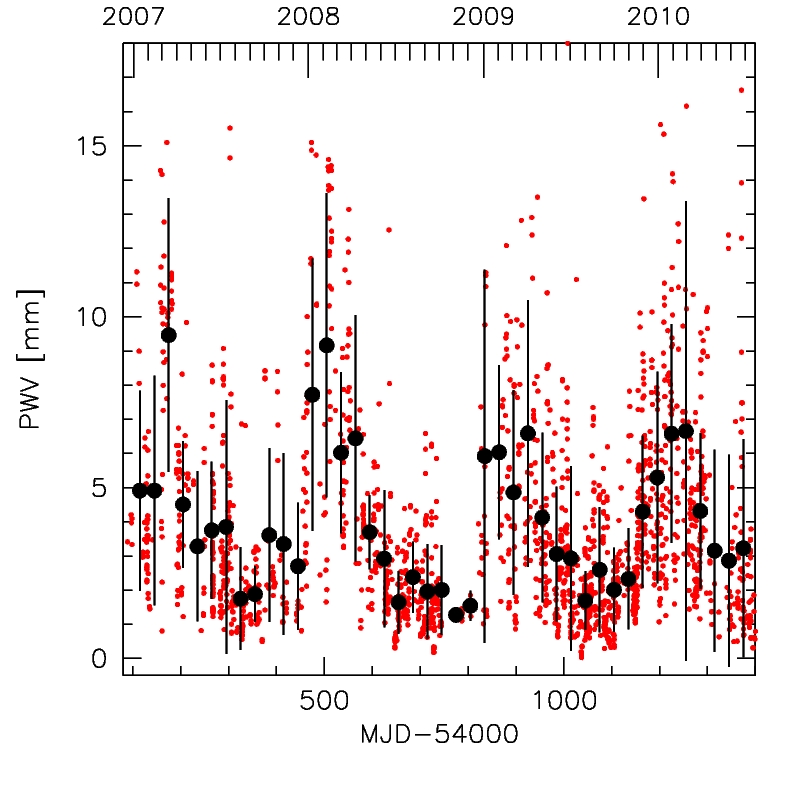}
}
\caption{\label{fig:pwvtrend}Precipitable water vapor measured in
  Paranal between 2007 and 2010 (Smette et al. \cite{smette}). The
  large dots and error bars indicate the monthly averages and the
  associated RMS deviations, respectively.}
\end{figure}


\subsubsection{Oxygen bands}

Molecular oxygen shows two main O$_2$ vibrational absorption bands
centred at 6870 \AA\/ and 7605\AA, usually indicated as B and A bands,
respectively (Fig.~\ref{fig:h2o}). Their typical equivalent widths
(EW) are $\sim$6\AA\/ and $\sim$28\AA, which make them easily
detectable in low resolution spectra. To quantify the variability of
O$_2$ column density, and its effect on the extinction, we have
measured the EW of the B band in the red setting spectra. The A band
is severely affected by fringing, and so no very accurate measurements
were possible. However, the integrated strengths of the two bands are
well correlated, as demonstrated by a series of LBLRTM simulations run
for different values of the column density N(O$_2$). Additionally,
they both follow a linear dependency on airmass, down to $X$=2.5.

The measurements clearly show the EW airmass dependency, which is well
reproduced by the following best fit relation: 

\begin{displaymath}
EW_{6870} = 5.47\pm0.04 + (2.56\pm0.06)\;(X-1)
\end{displaymath}

\noindent where $EW_{6870}$ is expressed in \AA. With the aid of this
relationship one can correct the observed values to zenith, and derive
the column density using a standard curve of growth procedure. For
this purpose we computed a number of LBLRTM models varying N(0$_2$)
between 8.4$\times$10$^{23}$ cm$^{-2}$ and 6.7$\times$10$^{24}$
cm$^{-2}$. Subsequently we measured the EW of the B band on the output
spectra, after convolving them with a Gaussian profile to simulate the
instrumental broadening (FWHM=12\AA). A best fit using a second order
polynomial gives the following result:

\begin{displaymath}
\log \mbox{N(O$_2$)} = 23.15 + 0.308\; EW_{6870} + 0.013\; EW^2_{6870}
\end{displaymath}

The zenith-corrected values of $EW_{6870}$ show a peak-to-peak range
of 2.5\AA, with an average value of 5.46\AA\/ (RMS=0.42\AA). This
corresponds to $\log \mbox{N(O$_2$)}$=24.4 cm$^{-2}$ (2.5$\times$10$^{24}$
cm$^{-2}$), which is close to the value predicted by the LBLRTM model
(24.52, or 3.3$\times$10$^{24}$ cm$^{-2}$). See also the match between
the model and real data in Fig.~\ref{fig:h2o}. Incidentally, as
proposed by Stevenson (\cite{stevenson}), synthetic absorption spectra
of this quality can be proficiently used to correct low resolution
data for telluric features.

To evaluate the impact of O$_2$ variability on the extinction, we have
run a series of LBLRTM models for different values of N(O$_2$) within
the range deduced from the data. Although the variation in EW is
significant (45\% peak-to-peak), the effect remains confined to
the two strong bands. The extinction coefficient variation is less
than 0.01 mag airmass$^{-1}$ in $R$ and $I$, while no measurable
effect is seen in $U$, $B$ and $V$ passbands. The total contribution
of O$_2$ to the broad-band extinction is 0.01 and 0.03 mag
airmass$^{-1}$ in $R$ and $I$, respectively, while it is below 0.002
in all other pass-bands\footnote{These values have been estimated
using standard passband curves, and assuming the spectrum of the
source to be flat within the relevant passband.}.

As far as the time evolution is concerned, the data show short
timescale variations (cf. Sect.~\ref{sec:varshort}), while there is
no statistically significant signature of a long term trend over the
six months spanned by our observations.  Also, a correlation analysis
between the EW of O$_2$ and the extinction coefficient derived on each
single spectrum in the red setting had shown that these are completely
independent ($r_{xy}<$0.2), irrespective of the wavelength under
consideration.

\subsubsection{Water bands}

In the optical there are three bands that contribute to the
extinction, at 7200, 8200 and 9400 \AA, the latter being the most
prominent one (Fig.~\ref{fig:trans}) and affecting the $z$
passband. The intensity of these absorption bands is well correlated
to the Precipitable Water Vapor (PWV), so that a best fit using an
appropriate radiation transfer model and suitable vertical profiles
enables the retrieval of the PWV amount directly from observed
high-resolution spectra (see for instance Smette, Horst \& Navarrete
\cite{smette}). A similar method can be used to roughly estimate the
PWV from the global EW of a given water band measured on low
resolution spectra. This can be then translated into H$_2$O column
density, and finally converted into PWV. An example is shown in
Fig.~\ref{fig:h2o}, which illustrates the very good match between the
data and an LBLRTM model with PWV=6mm. Using LBLRTM simulations run
with different values of water column density, we have derived the
following relation for the band at 7200 \AA, which holds for
PWV$\leq$10 mm:

\begin{displaymath}
PWV\approx 0.4\; EW_{7200} + 0.01 \; EW^2_{7200}
\end{displaymath}

\begin{figure}
\centerline{
\includegraphics[width=9cm]{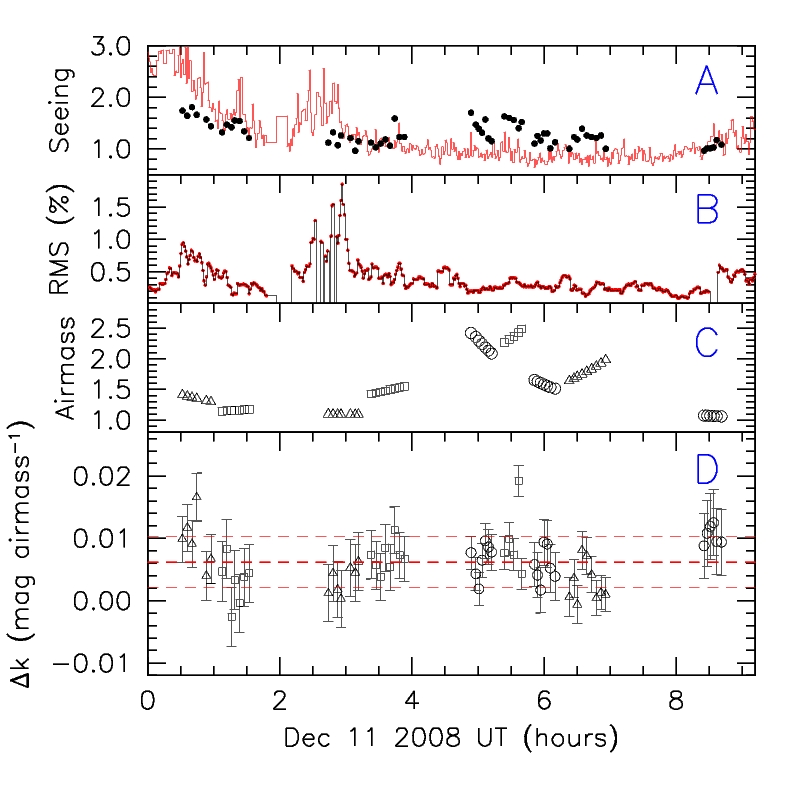}
}
\caption{\label{fig:varshort} Short timescale variations observed on
  Dec 11, 2008 at 5000 \AA. Panel A: DIMM seeing (the dots
  indicate the image quality measured on the spectra). Panel B: LOSSAM RMS
  fluctuations. Panel C airmass. Panel D: deviations from the average
  extinction at 5000 \AA. Different symbols refer to different stars: GD108
  (circles), GD50 (triangles), and BPM16274 (squares). The horizontal
  dashed lines trace the average deviation (middle) and $\pm$1 sigma
  levels (upper and lower).}
\end{figure}

Before entering the EW values into this relation, one needs to correct
to zenith the values measured at airmass $X$. This can be done through
the following formula, which was derived from a series of LBLRTM
simulations:

\begin{displaymath}
EW_0 = EW(X) - 3.59\; (X-1)
\end{displaymath}

In principle, provided that the signal-to-noise on the continuum is
sufficient ($>$50 per pixel), this relation can be used to retrieve
the amount of PWV from low resolution optical spectra.  For PWV=2mm,
which is a typical value for Paranal (see below), it is
$EW_{7200}$=4.1\AA. Unfortunately, due to the severe fringing, and the
relative weakness of the feature, we could not directly measure the
amount of PWV in our spectra. Therefore, to estimate the variability
of water features in the optical domain and its impact on extinction,
we used the PWV data obtained on Paranal (Smette et al. \cite{smette})
during 2008 and 2009. The PWV shows a strong seasonal dependency (see
Fig.~\ref{fig:pwvtrend}), with peak values exceeding 15mm in
February--March. The distribution has a median value of 2.6, and
in 95\% of the cases is PWV$<$10 mm. 

Variations in the amount of PWV have a mild effect on the extinction
coefficients, and this is limited to $R$ and $I$ passbands only. The
synthetic broad-band coefficients deduced from LBLRTM simulations
increase by 0.01 and 0.02 mag airmass$^{-1}$ in $R$ and $I$
respectively, if PWV changes from 2mm to 10mm.

\subsection{\label{sec:varshort}Short timescale variations}

The PARSEC data were collected with the aim of giving a statistically
robust description of Paranal extinction. Therefore, observations have
been spread as much as possible across the six months range covered by
the project. However, on a few occasions, the same standard stars were
observed at different airmasses during the same night, hence enabling
the the calculation of the specific night extinction, and the
study of its variability on the scales of tens of minutes.  The best
data set was obtained on Dec 11, 2008, and it contains a number of
repeated observations of GD108, GD50, and BPM16274. The data, spanning
almost the whole duration of the night, are presented in
Fig.~\ref{fig:varshort} for $\lambda$=5000 \AA.

The extinction (on average 0.006 mag airmass$^{-1}$ higher
than the best fit value 0.157$\pm$0.003), shows an RMS variation of
0.004 mag airmass$^{-1}$, which is comparable to the typical
measurement error. Judging from the DIMM intensity fluctuations
(Fig.~\ref{fig:varshort}, panel B), the night was
stable, with RMS flux fluctuations at 5000 \AA\/ below 1.5\% on the
scale of one minute. This is indeed reflected on the stability of the
extinction, which appears to vary by a similar amount. We also notice
that data taken at high airmass do not show a systematic bias towards
larger extinctions.

\section{\label{sec:dis}Discussion}

\begin{figure}
\centerline{
\includegraphics[width=9cm]{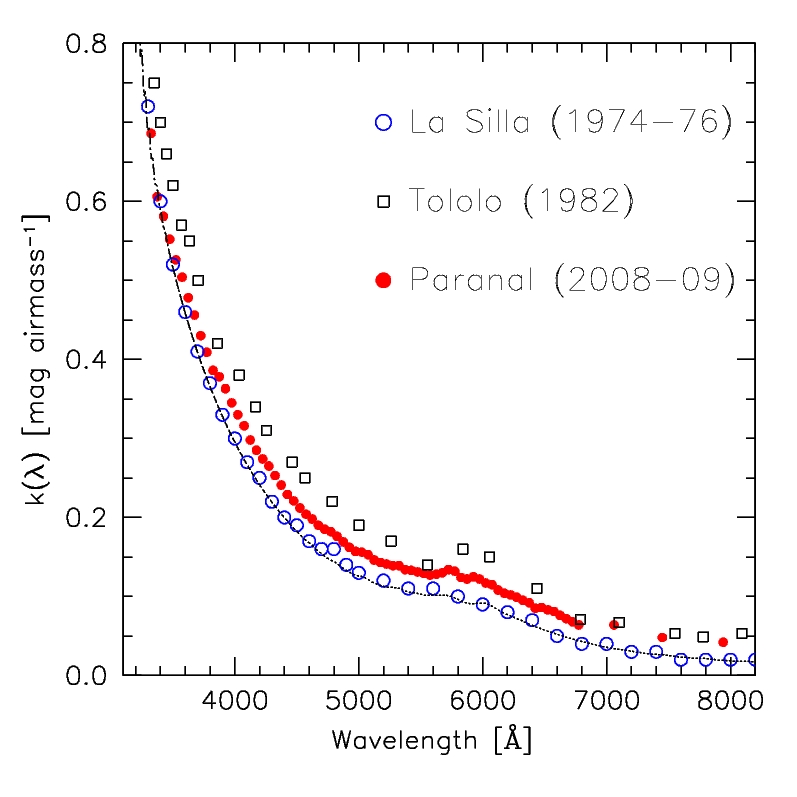}
}
\caption{\label{fig:comp} Comparison between the extinction curves of
  Paranal (this work), La Silla (T\"ug \cite{tug}), and Cerro Tololo
  (Stone \& Baldwin \cite{stone}). For comparison, the dashed curve
  traces a LBLRTM model computed for La Silla without aerosols.}
\end{figure}

The average extinction curve of Paranal, presented here for the first
time, appears to conform to the expectations for the site. However, it
is interesting to compare it to the data obtained for other two major
observatories in the Atacama desert, i.e. Cerro Tololo (2200 m a.s.l.)
and La Silla (2400 m a.s.l.), which are located within about 600 km
from Paranal. The comparison is presented in Fig.~\ref{fig:comp}. The
data for Cerro Tololo are taken from the atmospheric extinction file
{\tt ctioextinct.dat} included in IRAF (Stone \& Baldwin \cite{stone};
Baldwin \& Stone \cite{baldwin}). For La Silla we used the table
published in Schwarz \& Melnick (\cite{esomanual}), who reported data
obtained by T\"ug (\cite{tug}). These data are included in the {\tt
  atmoexan} table of the MIDAS distribution.

While Paranal and Cerro Tololo display similar behaviors (but see the
discussion below), La Silla shows a systematically lower extinction
with the maximum deviation ($\sim$0.05 mag airmass$^{-1}$) occurring
at about 4000 \AA.  Interestingly, the La Silla extinction curve is
very well matched by an LBLRTM simulation with no aerosols
(Fig.~\ref{fig:comp}, dotted line). These exceptionally low values were
noted already by T\"ug, who had derived a maximum aerosol contribution
of 0.01--0.02 mag airmass$^{-1}$ (see also Minniti et
al. \cite{minniti} for a comparison with two Argentinian sites). Since
the extinction curve for La Silla was derived using data collected
between 1974 and 1976, one possible explanation is that the
transparency conditions have degraded in the Atacama desert in the
last 35 years.

Although it must be noticed that the extinction curve of La Silla was
obtained selecting only the best nights (T\"ug \cite{tug}), the
increased opacity is most likely due to a systematic change in the
atmosphere above these sites. Already Stone \& Baldwin (\cite{stone})
had noticed this fact, and wrote that {\it these observations indicate
  a non-grey increase of the extinction at Cerro Tololo, compared to
  values in use half a decade ago, averaging about 0.04 mag per
  airmass in the red and about 0.08 mag per airmass in the UV}.  In
this respect it is important to note that the observations of Stone \&
Baldwin were most likely carried out before the volcanic eruption of El
Chich\'on (March, April 1982)\footnote{The authors do not report the
  actual dates of the observations, but their paper was submitted in
  June, 1982.}, which severely affected the transparency at CTIO and
La Silla (see Burki et al. \cite{burki} for a comprehensive analysis).

But the most intriguing aspect shown by this comparison is the fact
that while Paranal's curve is very similar to the one of CTIO above
6500 \AA, it gradually tends to the La Silla curve in the blue, so that
they attain very similar values below 3800 \AA\/
(Fig.~\ref{fig:comp}). The simplest explanation for this is a temporal
evolution of the aerosol mixture above the Atacama desert.

\begin{figure*}
\centerline{
\includegraphics[width=19cm]{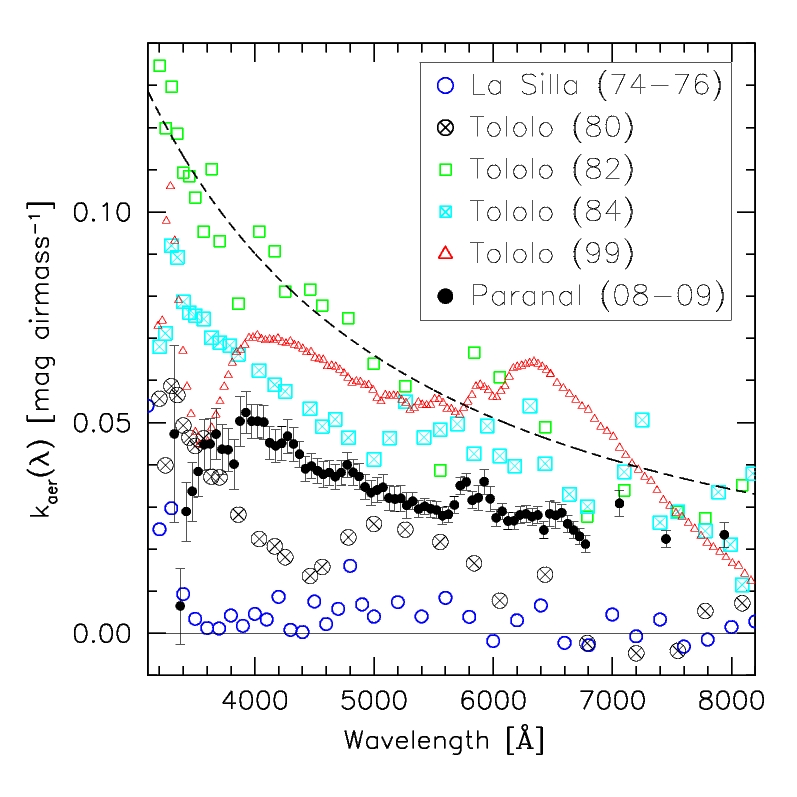}
}
\caption{\label{fig:comp2} Comparison between the aerosol extinction
  in La Silla (1974--1976: T\"ug \cite{tug}), Cerro Tololo (1980:
  Guti\'errez-Moreno et al. \cite{gutierrez82}; 1982: Stone \& Baldwin
  \cite{stone}; 1984: Guti\'errez-Moreno et al. \cite{gutierrez86};
  2004: Stritzinger et al. \cite{max}), and Cerro Paranal (this
  work). The dashed line traces the $k_{aer}=k_0 \lambda^{-\alpha}$
  best fit law to the Cerro Tololo 1982 data ($k_0$=0.025 mag
  airmass$^{-1}$, $\alpha$=$-$1.4).}
\end{figure*}

To test this hypothesis, we collected a number of published
spectroscopic extinction curves obtained on different epochs between
1974 and 2009. Then we computed the pure aerosol contribution as we
did in Sect.~\ref{sec:model}. The result is presented in
Fig.~\ref{fig:comp2}.  As expected, the 1974--1976 La Silla extinction
curve (T\"ug \cite{tug}) shows an aerosol contribution which is
smaller than 0.01 mag airmass$^{-1}$. Then, the data obtained at Cerro
Tololo in July--August 1980 (Guti\'errez-Moreno et
al. \cite{gutierrez82}) show a clear increase in the aerosol
extinction, especially in the blue domain, where it reaches 0.05 mag
airmass$^{-1}$. This increase becomes more marked in the next data
set, obtained at Cerro Tololo in 1982 (Stone \& Baldwin \cite{stone}),
as we said presumably before the El Chich\`on eruption. The aerosol
extinction is $\sim$0.07 mag airmass$^{-1}$ at 5000 \AA, and exceeds
0.10 mag airmass$^{-1}$ below 4000 \AA. The wavelength dependency is
well reproduced by a power law with $k_0$=0.025 mag airmass$^{-1}$,
and $\alpha=-$1.4 (see Sect.~\ref{sec:model}), as is typical of
atmospheric haze (Burki et al. \cite{burki}).

The next curve was obtained averaging the four data sets obtained at
Cerro Tololo on January, April, May and July 1984 (Guti\'errez-Moreno
et al. \cite{gutierrez86}), i.e. about 2 years after the El Chich\`on
eruption. Although the data above 6000 \AA\/ are a bit noisy, it is
clear that while the extinction in the red remained at levels
comparable to those measured in 1982, it decreased in the blue, making
the wavelength dependency significantly shallower. This is usually
interpreted as due to the evolution of the average particle size to
larger values (see for instance Schuster \& Parrao \cite{schuster}).

The subsequent major volcanic event was the eruption of Pinatubo,
which took place in June 1991. This produced very clear consequences
on the extinction (Grothues \& Gochermann \cite{grot}; Burki et al.
\cite{burki}; Schuster \& Parrao \cite{schuster}; Parrao \& Schuster
\cite{parrao}). Since then, a number of intermediate size events took
place, none reaching the level of the Pinatubo event.  The most
  relevant for our study is the explosive eruption of Chaiten
  (Southern Chile, May 2008). However, at variance with those of El
  Chich\'on and Pinatubo, Chaiten produced a very small amount of
  SO$_2$, which is the main contributor to aerosol pollution. For this
  reason we tend to exclude this eruption had major effects on the
  atmosphere above the Chilean astronomical sites. In this respect we
  note that an enhancement in the SO$_2$ content can be also produced
  by local volcanoes that are not erupting, but many of which have
  strong vapor vents.  These may have variations in intensity of
  activity that are not regularly monitored or observed.

The most recent extinction curve for the Atacama desert has been
published by Stritzinger et al. (\cite{max}), and it was obtained from
data taken at CTIO on seven nights in February 1999. The resulting
aerosol term shows a rather complex behavior, with a rapid decrease
above $\sim$6400 \AA, a rather flat plateau between 4000 and 6400 \AA,
and a drop followed by a rapid increase below 4000 \AA\/
(Fig.~\ref{fig:comp2}). Although some of the wiggles might be
generated by the procedure used to derive the curve, the overall
wavelength dependency is certainly different from the one expected for
a typical atmospheric haze.

The deviation from an ordinary aerosol law is confirmed by our data
set, which follows rather well (although with a shift of $\sim$0.01
mag airmass$^{-1}$) the points obtained at CTIO in 1984
(Guti\'errez-Moreno et al. \cite{gutierrez86}), with one remarkable
difference, i.e. the downturn below 4000 \AA. A similar drop has been
observed by Schuster \& Parrao (\cite{schuster}) in their data
obtained at S. Pedro M\'artir about one year after the Pinatubo
eruption. Small or even negative power indexes have also been reported
by Sterken \& Manfroid (\cite{sterken}) and Burki et
al. (\cite{burki}) as typical of volcanic pollutants. We finally note that
the curve by Stritzinger et al. (\cite{max}) also shows a downturn in
the blue, very similar to that displayed by the Paranal data. Parrao
\& Schuster (\cite{parrao}) have interpreted this UV drop as a
possible evidence for a masking (or decrease) of the normal
atmospheric opacity produced by the presence of volcanic aerosols (see
also the discussion in Sterken \& Manfroid \cite{sterken}). In
addition, this might also be the signature of a more complex, possibly
bi-modal, extinction law for the volcanic contribution (Parrao \&
Schuster \cite{parrao}).

What is surprising about these results is that such anomalous
wavelength dependencies are found almost twenty years after the
Pinatubo event, when one would expect the atmospheric transmittance to
be back to normal. But, as a matter of fact, the exceptional values
recorded in La Silla in the 1970s were never observed again. The data
collected in La Silla between 1974 and 1995 clearly show that in the
almost ten years that separated El Chich\`on and Pinatubo eruptions,
the $UBV$ extinction never reached the levels preceding the first
event (see Fig.~3 in Burki et al. \cite{burki}, and Schwartz
\cite{schwartz}). The extinction data presented in this work are on
average $\sim$0.01 mag airmass$^{-1}$ lower than those measured at
CTIO about two years after the El Chich\`on eruption
(Guti\'errez-Moreno et al. \cite{gutierrez86}), and $\sim$0.03 mag
airmass$^{-1}$ lower than those published by Stritzinger et
al. (\cite{max}) for the same observatory. This might be the signal
that the atmospheric transparency is slowly approaching the high
values observed prior to the great eruptions that took place at the
end of last century.

Further measurements in the next years will be needed to confirm
  the trend observed so far.

\section{\label{sec:conc} Conclusions}

In this paper we presented the best fit extinction curve for Cerro
Paranal, obtained combining spectroscopic data collected over more
than 40 nights between October 2008 and March 2009. The main results
of our analysis can be summarised as follows:

\begin{itemize} 

\item Above 4000\AA\/ the curve is well fitted by an atmospheric
  model in which the aerosol is described by a power law of the form
  $k_0\;\lambda^\alpha$. Below this wavelength the data show a
  systematic deficit in the extinction, which exceeds $\sim$0.03 mag
  airmass$^{-1}$ at 3700 \AA. Although this needs to be
    investigated with further observations, it may indicate the
  presence of pollutants of volcanic origin in the atmosphere above
  the Atacama desert.

\item During the six months covered by our observations, the
  extinction distribution is characterised by a semi-interquartile
  range of 0.01 mag airmass$^{-1}$ above 4000 \AA, with peak-to-peak
  variations up to $\sim$0.1 mag airmass$^{-1}$ in the UV. 

\item While the extinction measured at the various wavelengths is well
  correlated above 4000 \AA, the correlation is much weaker below this
  limit. This supports the conclusion that the UV portion of the
  aerosol contribution follows an independent behavior.

\item During the observing campaign, the Rayleigh scattering component
  was stable to $\sim$0.2\% (RMS). The much larger extinction
  fluctuations, seen at all wavelengths and reaching peak values
  exceeding 0.1 mag airmass$^{-1}$ are attributable to changes in the
  aerosol amount and (possibly) composition. The median value of the
  aerosol contribution at 4000 \AA\/ is 0.05 mag airmass$^{-1}$.

\item Ozone was found to vary by $\sim$5\% (RMS) around an average
  value of 258 DU, in good agreement with that derived from the OMI
  data (260 DU).
 
\item Although the O$_2$ A and B bands were found to vary by about
  45\%, the effect on the variation of broad-band $R$ and $I$
  extinction coefficients is smaller than 0.01 mag airmass$^{-1}$. The
  column density of O$_2$ derived from the data is fully consistent
  with the LBLRTM prediction. No seasonal effect was detected.

\item In the last 35 years the extinction above the Atacama desert has
  shown a marked evolution, most likely due to the two major eruptions
  of El Chich\'on and Pinatubo. The exceptionally low aerosol content
  measured in La Silla in 1974--76 was never re-established,
  indicating that it probably takes several decades before the
  pollutants completely fall out.

\item The usage of the IRAF extinction curve {\tt ctioextinct.dat}
  leads to systematic flux overestimates of more than 4\% below
  4000 \AA. Also, it tends to over-correct by $\sim$0.03 mag
  airmass$^{-1}$ at about 6000 \AA, which corresponds to the ozone
  bump. On the other hand, the usage of the {\tt atmoexan} MIDAS table
  produces systematic flux underestimates that amount to $\sim$2\% at
  8000 \AA, and exceed 5\% at 4000 \AA.

\item The adoption of the best fit curve presented in this paper for
  the extinction correction of spectroscopic data obtained at Paranal
  under clear-sky conditions leads to an RMS uncertainty of about 1\%.

\end{itemize}


\begin{acknowledgements} 
This paper is based on calibration data obtained with the ESO Very
Large Telescope at Paranal Observatory. We are indebted to W. Kausch
and M. Barden for the very useful help received during the setup of
the LBLRTM code. The support of C. Izzo for the usage of the FORS
pipeline is kindly acknowledged. We are grateful to A. Kaufer and
R. Fosbury for reading and commenting the manuscript. Special
  thanks go to C.R. Stern and J. Mu\~noz for the information on the
  Chaiten eruption. We finally wish to thank the referee, C. Sterken,
  for his careful manuscript review and valuable suggestions.
\end{acknowledgements}

%
%
\Online

\appendix

\section{\label{sec:seeing} Correction for slit losses due to seeing}

\begin{figure}
\centerline{
\includegraphics[width=9cm]{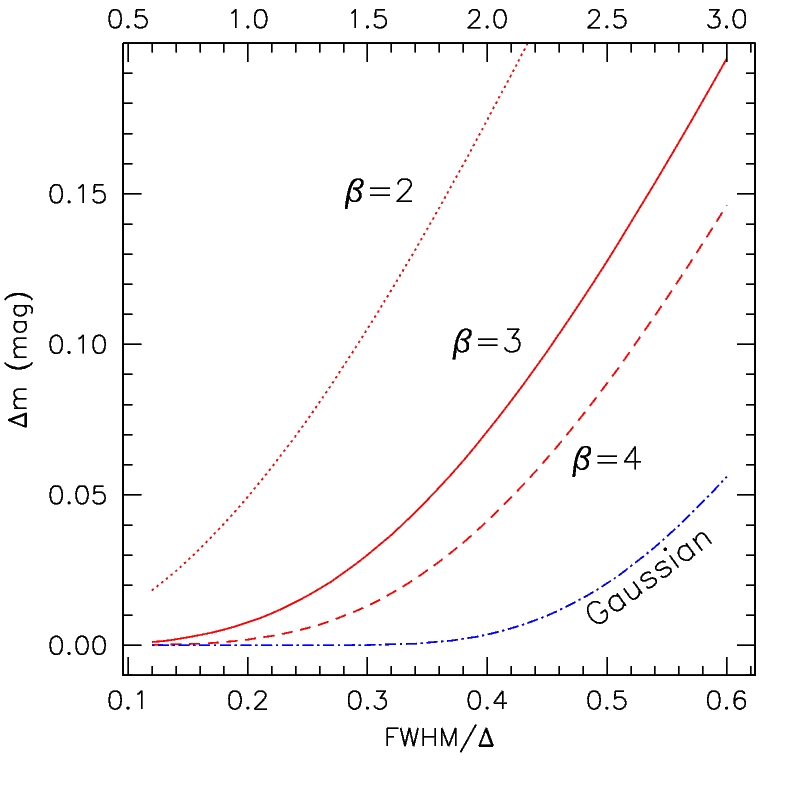}
}
\caption{\label{fig:seeing} Slit losses for Moffat PSFs with different
  $\beta$ values. For comparison, the Gaussian case is also plotted
  (dotted-dashed line). The upper scale indicates the FWHM seeing (in
  arcsec) for a slit width $\Delta$=5 arcsec.}
\end{figure}

Because of the extended wings of the point spread function (PSF)
typically delivered by telescopes, and depending on the slit width
$\Delta$ used for the observations, a fraction of the incoming flux
does not enter the spectrograph. If this loss were constant, there
would be no effect on the final estimate of the extinction
coefficients. However, data are obtained under different seeing
conditions, and seeing tends to be larger at larger
airmasses. Therefore, this introduces a systematic effect which leads
to artificially higher extinction estimates. Obviously, these losses
cannot be reduced by placing the slit along the parallactic angle or
observing through an atmospheric dispersion corrector.

In this section we describe the method we have used to correct the
instrumental magnitudes derived from our spectra. For this purpose,
let us introduce a coordinates reference system placed on the focal
plane of the telescope, with its origin on the optical axys. Then let
us indicate with $P(x,y)$ the PSF, normalised so that:

\begin{displaymath}
 \int_{-\infty}^{+\infty}  \int_{-\infty}^{+\infty} P(x,y) \; dx \; dy = 1
\end{displaymath}

If the slit width is $\Delta$, then the slit
flux losses (expressed in magnitudes) can be computed as:

\begin{equation}
\label{eq:loss}
\Delta m = -{\mbox2.5}\log \int_{-\infty}^{+\infty}  \int_{-\Delta/2}^{+\Delta/2}
 P(x,y) \; dx \; dy
\end{equation}

\noindent for any PSF profile. For our calculations we adopted the
profile derived by Moffat (\cite{moffat}):

\begin{displaymath}
P(x,y)=\left [  1 + \frac{x^2 + y^2}{\sigma^2} \right ]^{-\beta}
\end{displaymath}

\noindent where the width parameter $\sigma$ is related to the FWHM as
follows:

\begin{displaymath}
\sigma=\frac{FWHM}{2\sqrt{2^{1/\beta}-1}}
\end{displaymath}

\begin{figure}
\centerline{
\includegraphics[width=9cm]{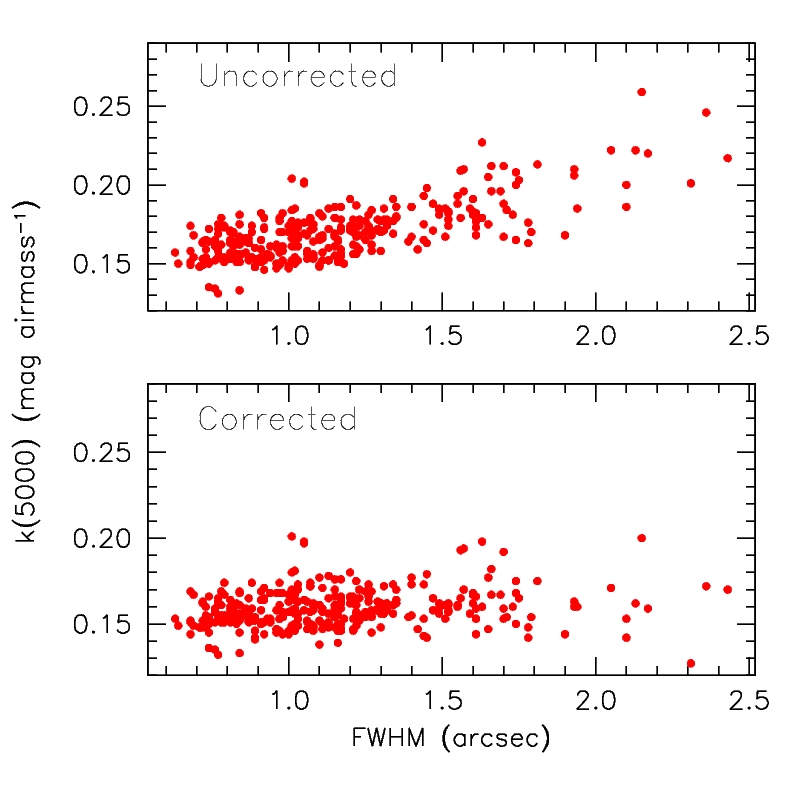}
}
\caption{\label{fig:seeingcorr} Effect of the slit losses correction
  on the PARSEC extinction at 5000 \AA. Upper panel: original
  data. Lower panel: corrected data.}
\end{figure}

The typical values of $\beta$, deduced from observed stellar profiles,
range between 2.5 and 4.0 (Saglia et al. \cite{saglia}). For our
calculations we have used $\beta$=3, which gives a very good fit to
the VLT FORS1 data in the wavelength range of interest
(3300--8000 \AA). Equation~\ref{eq:loss} can be integrated numerically
for different values of the FWHM, and the correction $\Delta m$
readily derived. A real example is illustrated in
Fig.~\ref{fig:seeing}. For a slit width of 5 arcsec (as is our case),
with a seeing of 1 arcsec the losses amount only to $\sim$0.008 mag,
but they grow to $\sim$0.07 mag for a seeing of 2 arcsec.

The effect of the correction is illustrated in
Fig.~\ref{fig:seeingcorr}, which presents the PARSEC extinction data
at 5000 \AA. The uncorrected data show a clear dependency of $k$ from
the seeing (the Pearson correlation factor is $r_{xy}$=0.71), while
the corrected values show no correlation ($r_{xy}$=0.21). Since the
seeing grows as $\lambda^{-1/5}$ (Roddier \cite{roddier}), slit losses
are higher in the blue than in the red. Therefore, not only does one
expect a correlation between the uncorrected instrumental magnitudes
and the seeing, but also that this relation becomes steeper in the
blue. This expectation is confirmed by our data. An inspection of the
PARSEC data at various wavelengths shows that the application of the
correction removes any dependency of $k$ from seeing and airmass. As a
side effect, it also reduces the spread of the data points.

\section{\label{sec:tab} Tabulated extinction curve for Paranal}

\begin{table*}
\caption{\label{tab:curve}Best fit extinction curve for Paranal}
\tabcolsep 5mm
\centerline{
\begin{tabular}{ccc|ccc|ccc}
\hline
$\lambda$&$k(\lambda)$&$\sigma_k$&$\lambda$&$k(\lambda)$&$\sigma_k$&$\lambda$&$k(\lambda)$&$\sigma_k$\\
\hline
  3325 &  0.686 &  0.021 &  4625 &  0.198 &  0.003 &  5925 &  0.125 &  0.003 \\
  3375 &  0.606 &  0.009 &  4675 &  0.190 &  0.003 &  5975 &  0.122 &  0.003 \\
  3425 &  0.581 &  0.007 &  4725 &  0.185 &  0.003 &  6025 &  0.117 &  0.002 \\
  3475 &  0.552 &  0.007 &  4775 &  0.182 &  0.003 &  6075 &  0.115 &  0.002 \\
  3525 &  0.526 &  0.006 &  4825 &  0.176 &  0.003 &  6125 &  0.108 &  0.002 \\
  3575 &  0.504 &  0.006 &  4875 &  0.169 &  0.003 &  6175 &  0.104 &  0.002 \\
  3625 &  0.478 &  0.006 &  4925 &  0.162 &  0.003 &  6225 &  0.102 &  0.002 \\
  3675 &  0.456 &  0.006 &  4975 &  0.157 &  0.003 &  6275 &  0.099 &  0.002 \\
  3725 &  0.430 &  0.006 &  5025 &  0.156 &  0.003 &  6325 &  0.095 &  0.002 \\
  3775 &  0.409 &  0.005 &  5075 &  0.153 &  0.003 &  6375 &  0.092 &  0.002 \\
  3825 &  0.386 &  0.006 &  5125 &  0.146 &  0.003 &  6425 &  0.085 &  0.002 \\
  3875 &  0.378 &  0.006 &  5175 &  0.143 &  0.003 &  6475 &  0.086 &  0.003 \\
  3925 &  0.363 &  0.005 &  5225 &  0.141 &  0.003 &  6525 &  0.083 &  0.003 \\
  3975 &  0.345 &  0.004 &  5275 &  0.139 &  0.003 &  6575 &  0.081 &  0.002 \\
  4025 &  0.330 &  0.004 &  5325 &  0.139 &  0.002 &  6625 &  0.076 &  0.002 \\
  4075 &  0.316 &  0.004 &  5375 &  0.134 &  0.002 &  6675 &  0.072 &  0.002 \\
  4125 &  0.298 &  0.004 &  5425 &  0.133 &  0.002 &  6725 &  0.068 &  0.002 \\
  4175 &  0.285 &  0.004 &  5475 &  0.131 &  0.002 &  6775 &  0.064 &  0.002 \\
  4225 &  0.274 &  0.004 &  5525 &  0.129 &  0.002 &  7060 &  0.064 &  0.003 \\
  4275 &  0.265 &  0.004 &  5575 &  0.127 &  0.002 &  7450 &  0.048 &  0.002 \\
  4325 &  0.253 &  0.004 &  5625 &  0.128 &  0.002 &  7940 &  0.042 &  0.003 \\
  4375 &  0.241 &  0.003 &  5675 &  0.130 &  0.002 &       &        &        \\
  4425 &  0.229 &  0.003 &  5725 &  0.134 &  0.002 &  8500 &  0.032 &  (*) \\
  4475 &  0.221 &  0.003 &  5775 &  0.132 &  0.002 &  8675 &  0.030 &  (*) \\
  4525 &  0.212 &  0.003 &  5825 &  0.124 &  0.002 &  8850 &  0.029 &  (*) \\
  4575 &  0.204 &  0.003 &  5875 &  0.122 &  0.003 & 10000 &  0.022 &  (*) \\
\hline
\multicolumn{9}{l}{(*) The last four values are interpolations from the LBLRTM model.}
\end{tabular}
}
\end{table*}

The merged extinction curve of Paranal is presented in
Table~\ref{tab:curve}. To allow the usage of the data across the whole
optical domain, we have added four points obtained from the
interpolation of an LBLRTM model. The wavelengths (8500, 8675, 8850,
and 10000 \AA) were selected to avoid the strong water absorption
bands. The LBLRTM simulation was run disabling the aerosol
calculation. Then we added the aerosol contribution using the best fit
relation $k_{\rm aer}=k_0 \lambda^\alpha$, with $k_0$=0.013 and
$\alpha$=$-$1.38 (see Sect.~\ref{sec:aer}).

\end{document}